\documentclass[twocolumn]{aastex631}
\usepackage{makecell}
\usepackage{natbib}
\usepackage{color}
\usepackage{mathptmx}
\usepackage{amsmath}
\usepackage{url}
\usepackage{multirow}
\usepackage{verbatim}
\usepackage{numprint}
\usepackage{graphicx}
\usepackage{refcount}
\usepackage{hyperref}

\usepackage[utf8]{inputenc}   
\usepackage[T1,T5]{fontenc}

\newcommand{\msun}{\,{\rm M}_\odot}
\newcommand{\pc}{\,{\rm pc}}
\newcommand{\kpc}{\,{\rm kpc}}
\newcommand{\Myr}{\,{\rm Myr}}
\newcommand{\yr}{\,{\rm yr}}

\definecolor{mycol}{rgb}{0.05, 0.4, 0.1}
\definecolor{mycol2}{rgb}{1, 0.05, 0.05}
\definecolor{mycol3}{rgb}{0.8, 1, 0.6}

\def\arxiv{1}   
\ifdefined\revision
  \def\my#1{{\textcolor{mycol2}{#1}}}
  \def\del#1{{\textcolor{mycol3}{#1}}}
\fi
\ifdefined\arxiv
  \def\my#1{#1}
  \def\del#1{}
\fi

\received{March XX, 2025}
\revised{March XX, 2025}
\accepted{April XX, 2025}

\shorttitle{{\it AGORA} Comparison. VIII: Disk Formation and Evolution of Milky Way Progenitors}
\shortauthors{{\it AGORA} Collaboration et al.}

\begin{document}
\title{The {\it AGORA} High-resolution Galaxy Simulations Comparison Project. VIII: Disk Formation and Evolution of Simulated Milky Way Mass Galaxy Progenitors at $1<z<5$}

\correspondingauthor{Minyong Jung, Ji-hoon Kim, Thịnh Hữu Nguyễn, Ram\'{o}n Rodr\'{i}guez-Cardoso}
\affiliation{\rm The authors marked with {}\textsuperscript{*} as code leaders contributed to the article by leading the effort within each code group to perform and analyze simulations.} 

\author[0000-0002-9144-1383]{Minyong Jung}
\affiliation{Center for Theoretical Physics, Department of Physics and Astronomy, Seoul National University, Seoul 08826, Republic of Korea;
\rm{\href{mailto:wispedia@snu.ac.kr}{wispedia@snu.ac.kr}}}

\author[0000-0003-4464-1160]{Ji-hoon Kim}
\altaffiliation{Code leaders}
\affiliation{Center for Theoretical Physics, Department of Physics and Astronomy, Seoul National University, Seoul 08826, Republic of Korea;
\rm{\href{mailto:mornkr@snu.ac.kr}{mornkr@snu.ac.kr}}}
\affiliation{Institute for Data Innovation in Science, Seoul National University, Seoul 08826, Republic of Korea}
\affiliation{Seoul National University Astronomy Research Center, Seoul 08826, Republic of Korea}

\author[0009-0002-2290-8039]{Thịnh Hữu Nguyễn}
\affil{Department of Astronomy, University of Illinois at Urbana-Champaign, Urbana, IL 61801, USA; \rm{\href{mailto:thinhhn2@illinois.edu}{thinhhn2@illinois.edu}}}
\affil{Center for AstroPhysical Surveys, National Center for Supercomputing Applications, Urbana, IL, 61801, USA}

\author[0000-0002-9158-195X]{Ram\'{o}n Rodr\'{i}guez-Cardoso}
\affil{Departamento de Física de la Tierra y Astrofísica, Fac. de C.C. Físicas, Universidad Complutense de Madrid, E-28040 Madrid, Spain;
\rm{\href{mailto:ramorodr@ucm.es}{ramorodr@ucm.es}}}
\affil{GMV, Space and Avionics Equipment, Isaac Newton, 11 Tres Cantos, E-28760 Madrid, Spain}
\affil{Instituto de Física de Partículas y del Cosmos, IPARCOS, Fac. C.C. Físicas, Universidad Complutense de Madrid, E-28040 Madrid, Spain}

\author[0000-0002-6299-152X]{Santi Roca-F\`{a}brega}
\altaffiliation{Code leaders}
\affil{Lund Observatory, Division of Astrophysics, Department of Physics, Lund University, SE-221 00 Lund, Sweden}
\affil{Departamento de F\'{i}sica de la Tierra y Astrof\'{i}sica, Facultad de Ciencias F\'{i}sicas, Plaza Ciencias, 1, 28040 Madrid, Spain}

\author[0000-0001-5091-5098]{Joel R. Primack}
\affil{Department of Physics, University of California at Santa Cruz, Santa Cruz, CA 95064, USA}

\author[0000-0002-8638-1697]{Kirk S.~S.~Barrow}
\affiliation{Department of Astronomy, University of Illinois at Urbana-Champaign, Urbana, IL 61801, USA}

\author[0000-0003-0073-3012]{Anna Genina}
\altaffiliation{Code leaders}
\affil{Max-Planck-Institut f\"{u}r Astrophysik, Karl-Schwarzschild-Str. 1, D-85748, Garching, Germany}

\author[0009-0002-1398-6537]{Pablo Granizo}
\affiliation{Theoretical Astrophysics, Department of Earth and Space Science, Graduate School of Science, Osaka University, Toyonaka, Osaka, 560-0043, Japan}
\affil{Universidad Aut\'{o}noma de Madrid, Ciudad Universitaria de Cantoblanco, E-28049 Madrid, Spain}

\author[0000-0002-7820-2281]{Hyeonyong Kim}
\altaffiliation{Code leaders}
\affiliation{Center for Theoretical Physics, Department of Physics and Astronomy, Seoul National University, Seoul 08826, Korea}

\author[0000-0001-7457-8487]{Kentaro Nagamine}
\altaffiliation{Code leaders}
\affiliation{Theoretical Astrophysics, Department of Earth and Space Science, Graduate School of Science, Osaka University, Toyonaka, Osaka, 560-0043, Japan}
\affiliation{Theoretical Joint Research, Forefront Research Center, Graduate School of Science, Osaka University, Toyonaka, Osaka 560-0043, Japan}
\affiliation{Kavli IPMU (WPI), University of Tokyo, 5-1-5 Kashiwanoha, Kashiwa, Chiba, 277-8583, Japan}
\affiliation{Department of Physics \& Astronomy, University of Nevada Las Vegas, Las Vegas, NV 89154, USA}
\affiliation{Nevada Center for Astrophysics, University of Nevada, Las Vegas, 4505 S. Maryland Pkwy, Las Vegas, NV 89154-4002, USA} 

\author[0000-0002-5712-6865]{Yuri Oku}
\affil{Center for Cosmology and Computational Astrophysics, Institute for Advanced Study in Physics, Zhejiang University, Hangzhou 310027, People's Republic of China}

\author[0000-0002-3764-2395]{Johnny W. Powell}
\altaffiliation{Code leaders}
\affil{Department of Physics, Reed College, Portland, OR 97202, USA}

\author[0000-0002-6227-0108]{Yves Revaz}
\altaffiliation{Code leaders}
\affil{Institute of Physics, Laboratoire d'Astrophysique, \'{E}cole Polytechnique F\'{e}d\'{e}rale de Lausanne (EPFL), CH-1015 Lausanne, Switzerland}

\author{H\'{e}ctor Vel\'{a}zquez}
\altaffiliation{Code leaders}
\affil{Instituto de Astronom\'{i}a, Universidad Nacional Aut\'{o}noma de M\'{e}xico, A.P. 70-264, 04510, Mexico, D.F., Mexico}

\author[0000-0001-6106-7821]{Alessandro Lupi}
\altaffiliation{Code leaders}
\affil{DiSAT, Universit\`a degli Studi dell'Insubria, via Valleggio 11, I-22100 Como, Italy}
\affil{Dipartimento di Fisica ``G. Occhialini'', Universit\`a degli Studi di Milano-Bicocca, I-20126 Milano, Italy}

\author{Ikkoh Shimizu}
\altaffiliation{Code leaders}
\affil{Shikoku Gakuin University, 3-2-1 Bunkyocho, Zentsuji, Kagawa, 765-8505, Japan}

\author[0000-0002-5969-1251]{Tom Abel}
\affil{Kavli Institute for Particle Astrophysics and Cosmology, Stanford University, Stanford, CA 94305, USA}
\affil{Department of Physics, Stanford University, Stanford, CA 94305, USA}
\affil{SLAC National Accelerator Laboratory, Menlo Park, CA 94025, USA}

\author[0000-0002-4287-1088]{Oscar Agertz}
\affil{Lund Observatory, Division of Astrophysics, Department of Physics, Lund University, SE-221 00 Lund, Sweden}

\author[0000-0001-8531-9536]{Renyue Cen}
\affil{Center for Cosmology and Computational Astrophysics, Institute for Advanced Study in Physics, Zhejiang University, Hangzhou 310027, People's Republic of China}
\affil{Institute of Astronomy, School of Physics, Zhejiang University, Hangzhou 310027, People's Republic of China}

\author[0000-0002-8680-248X]{Daniel Ceverino}
\affil{Universidad Aut\'{o}noma de Madrid, Ciudad Universitaria de Cantoblanco, E-28049 Madrid, Spain}
\affil{CIAFF, Facultad de Ciencias, Universidad Aut\'{o}noma de Madrid, E-28049 Madrid, Spain}

\author[0000-0003-4174-0374]{Avishai Dekel}
\affil{Center for Astrophysics and Planetary Science, Racah Institute of Physics, The Hebrew University, Jerusalem 91904, Israel}

\author[0009-0009-6888-7967]{Chaerin Jeong}
\affil{Department of Astronomy \& Space Science, Kyung Hee University, 1732 Deogyeong-daero, Yongin-si, Gyeonggi-do 17104, Republic of Korea}

\author{Lucio Mayer}
\affil{Department of Astrophysics, University of Zurich, Winterthurerstrasse 190, CH-8057 Zürich, Switzerland}

\author[0000-0003-4597-6739]{Boon Kiat Oh}
\affiliation{Department of Physics, University of Connecticut, U-3046, Storrs, CT 06269, USA}

\author[0000-0001-5510-2803]{Thomas R. Quinn}
\affil{Department of Astronomy, University of Washington, Seattle, WA 98195, USA}


\author[0000-0002-4362-4070]{Hyunmi Song}
\affil{Department of Astronomy and Space Science, Chungnam National University, Daejeon 34134, Republic of Korea}


\author{for the {\it AGORA} Collaboration}
\affiliation{\rm \url{http://www.AGORAsimulations.org}}

\begin{abstract}
We investigate how differences in the stellar feedback produce disks with different morphologies in Milky Way-like progenitors over 1 $\leq z \leq 5$, using eight state-of-the-art cosmological hydrodynamics simulation codes in the \textit{AGORA} project. In three of the participating codes, a distinct, rotation-dominated inner core emerges with a formation timescale of $\lesssim 300 \Myr$, largely driven by a major merger event, while two other codes exhibit similar signs of wet compaction---gaseous shrinkage into a compact starburst phase---at earlier epochs. The remaining three codes show only weak evidence of wet compaction. 
Consequently, we divide the simulated galaxies into two groups: those with strong compaction signatures and those with weaker ones. 
Galaxies in these two groups differ in size, stellar age gradients, and disk-to-total mass ratios. 
Specifically, codes with strong wet compaction build their outer disks in an inside-out fashion, leading to negative age gradients, whereas codes with weaker compaction feature flat or positive age gradients caused primarily by outward stellar migration. 
Although the stellar half-mass radii of these two groups diverge at $z \sim 3$, the inclusion of dust extinction brings their sizes and shapes in mock observations closer to each other and to observed galaxies. 
We attribute the observed morphological differences primarily to variations in the stellar feedback implementations --- such as delayed cooling timescales, and feedback strengths --- that regulate both the onset and duration of compaction. 
Overall, our results suggest that disk assembly at high redshifts is highly sensitive to the details of the stellar feedback prescriptions in simulations. 

\end{abstract}


\section{Introduction} \label{sec:intro}

Disk galaxies form the most prevalent morphological class among galaxies with masses similar to the Milky Way (MW). 
The formation of galactic disks is theoretically understood as a result of the conservation of angular momentum of cooling gas within dark matter halos, leading the gas to settle into rotationally supported structures \citep[e.g.,][]{1980MNRAS.193..189F, 1998MNRAS.295..319M}. 
Numerical realizations of structural properties like thin disks in MW-mass galaxies have been achieved in cosmological contexts \citep[e.g.,][]{2011ApJ...728...51B, 2011ApJ...742...76G, 2017MNRAS.467..179G, 2018MNRAS.480..800H, 2021MNRAS.503.5826A, 2023MNRAS.522.3831F}, enabling studies on how the MW and its analogs form and grow their structures. 

Many theoretical models predict that galactic disks grow primarily from the inside out \citep[e.g.,][]{1998ApJ...507..601V, 2003ApJ...597...21A, 2014MNRAS.441.3679A, 2024ARNPS..74..173P}, in which new stars form at increasingly larger radii over time. Observations of disk galaxies support this picture: younger stellar populations often appear farther out, creating negative radial age gradients \citep[e.g.,][]{2014A&A...562A..47G, 2017MNRAS.466.4731G, 2019ApJ...884...99F}. However, smaller-radius star-forming galaxies at low redshifts often have more central star formation \citep{2024MNRAS.532.4217S}. \cite{2024arXiv241203455J} show that both negative and positive gradients exist in $z>4$ galaxies. \cite{2024OJAp....7E.113J} find that at $0.5 \lesssim z \lesssim 2$ galaxies above (below) the MS grow their centers (outskirts) first. 

Despite this progress, the mechanisms of disk formation at high redshift remain debated. Observations of disk galaxies at $z\sim2$ reveal thick and highly turbulent disks with higher star formation rates \citep{2013PASA...30...56G, 2018MNRAS.474.5076J, 2019ApJ...886..124W, 2019ApJ...880...48U}. However, it was unclear whether these disks evolve into the thick disks of present-day spirals, S0 galaxies, or massive ellipticals through major mergers \citep{2013PASA...30...56G}.

Recent observations from JWST have significantly increased the number of galaxy detections at high redshift, suggesting a high fraction of undisturbed disk-like morphologies \citep{2022ApJ...938L...2F, 2023ApJ...946L..15K, 2023ApJ...942L..42R, 2024A&A...685A..48H}. The large number of newly detected galaxies enables statistical studies to trace the temporal evolution of MW progenitors at $z>2$, indicating that Milky Way analog progenitors began forming disks and growing in an inside-out manner as early as $z\sim5$ \citep{2024arXiv241207829T}. JWST observations have also revealed that $z>1$ galaxies with $\log{M_{\star}/\msun} = 9-9.5$ are predominantly characterized by prolate morphologies \citep{2024ApJ...963...54P}.

Moreover, observations from the Atacama Large Millimeter/submillimeter Array (ALMA) have revealed that many recently discovered high-redshift disks are surprisingly well-settled, with rotation clearly dominating over random turbulent motions \citep[e.g.,][]{2023A&A...679A.129R, 2024A&A...689A.273R, 2024MNRAS.535.2068R}. Remarkably, some even exhibit distinct features such as spiral arms and bars \citep{2023ApJ...945L..10G, 2024ApJ...968L..15K}.

One channel for the formation of disk structures at high redshift involves a rapid compaction event induced by gas shrinkage. This process occurs as gas loses angular momentum through mergers or inflows of counter-rotating gas --- referred to as \textit{`wet compaction'} \citep{2014MNRAS.438.1870D, 2015MNRAS.450.2327Z, 2016MNRAS.458.4477T, 2023MNRAS.522.4515L}. The resulting inner mass condensation, accompanied by a starburst in the central region, is referred to as the \textit{blue nugget} phase. Following compaction, inside-out quenching and the subsequent formation of thin disks are expected.

\my{On the other hand, \cite{2023MNRAS.525.2241H} highlight a sufficiently concentrated mass profile as the primary driver of disk formation and the cessation of bursty star formation. The formation of bulges through inner mass concentration promotes the growth of thin disks \citep{2024ApJ...972...73S}.}

At high redshift, misaligned gas accretion triggers central star formation that may lead to the development of a spheroidal component or further central mass growth \citep{2012MNRAS.423.1544S, 2014MNRAS.441.3679A, 2020MNRAS.497.4346K}. Gas-rich galaxies---especially prevalent at cosmic noon---often undergo gas fragmentation and clump formation due to disk instabilities \citep{2009ApJ...703..785D, 2010MNRAS.404.2151C, 2010ApJ...719.1230A,2023MNRAS.519.6222D}, substantially contributing to the population of starburst galaxies at $z < 4$ \citep{2024arXiv240509619F}. These processes introduce significant diversity in the morphologies of high-redshift disk galaxies, deviating from purely inside-out growth and underscoring the intricate interplay between stellar feedback and gaseous disk kinematics.


Since galaxy morphology is tightly correlated with stellar feedback \citep{2005MNRAS.363.1299O, 2011MNRAS.415.1051B, 2015ApJ...804...18A, 2023MNRAS.522.3912C}, variations in subgrid feedback implementations can lead to significant differences in simulated galaxy shapes. Comparing results across different feedback schemes in a system with an identical merger history provides a broader understanding of disk formation mechanisms in the early universe.

The \textit{AGORA} simulation comparison project aims to enhance the realism and predictive power of galaxy simulations and deepen our understanding of the feedback processes that govern a galaxy's ``metabolism" \citep[][hereafter Papers I and II]{2014ApJS..210...14K, 2016ApJ...833..202K}. To achieve this goal, it compares a suite of simulations from multiple codes that share common initial conditions and astrophysics packages. Ultimately, the \textit{AGORA} project seeks to resolve long-standing challenges in galaxy formation and ensure that any successes arise from robust astrophysical assumptions rather than artifacts unique to specific codes.

In this study, we explore how baryonic processes influence disk formation by comparing cosmological zoom-in simulations with identical initial conditions of Milky Way-like galaxies across eight numerical codes from the \textit{AGORA} \textit{CosmoRun} \citep[][hereafter Paper III]{2021ApJ...917...64R}. Although these simulations exhibit good agreement in stellar masses, within 0.5--1.0 dex \citep[][hereafter Paper IV]{2024ApJ...968..125R}, and in the number and properties of satellite galaxies \citep[][hereafter Paper V]{2024ApJ...964..123J}, significant differences emerge in circumgalactic medium properties such as metal distribution and ionization states \citep[][hereafter Paper VI]{2024ApJ...962...29S}, as well as in satellite quenching \citep[][Paper VII]{2025arXiv250505844R}. Paper IV also reported notable disparities in stellar surface density profiles, Sérsic indices, and galaxy sizes across codes. 

This paper is organized as follows. In Section \ref{sec:method}, we summarize the \textit{CosmoRun} model and describe methods for analyzing stellar disk formation, generating mock observations, and measuring morphology. In Section \ref{sec:result}, we present the evolution of disk properties from $z=5$ to $z=1$, compare them with observational data, and examine the spatial distribution of stellar ages. In Section \ref{sec:discuss}, we discuss inter-code differences in the signatures and timing of wet compaction. Finally, in Section \ref{sec:conclusion}, we summarize our findings and provide concluding remarks.

\section{Method}\label{sec:method}
\begin{figure*}
    \centering
    \vspace{3mm}
    \includegraphics[width = \linewidth]{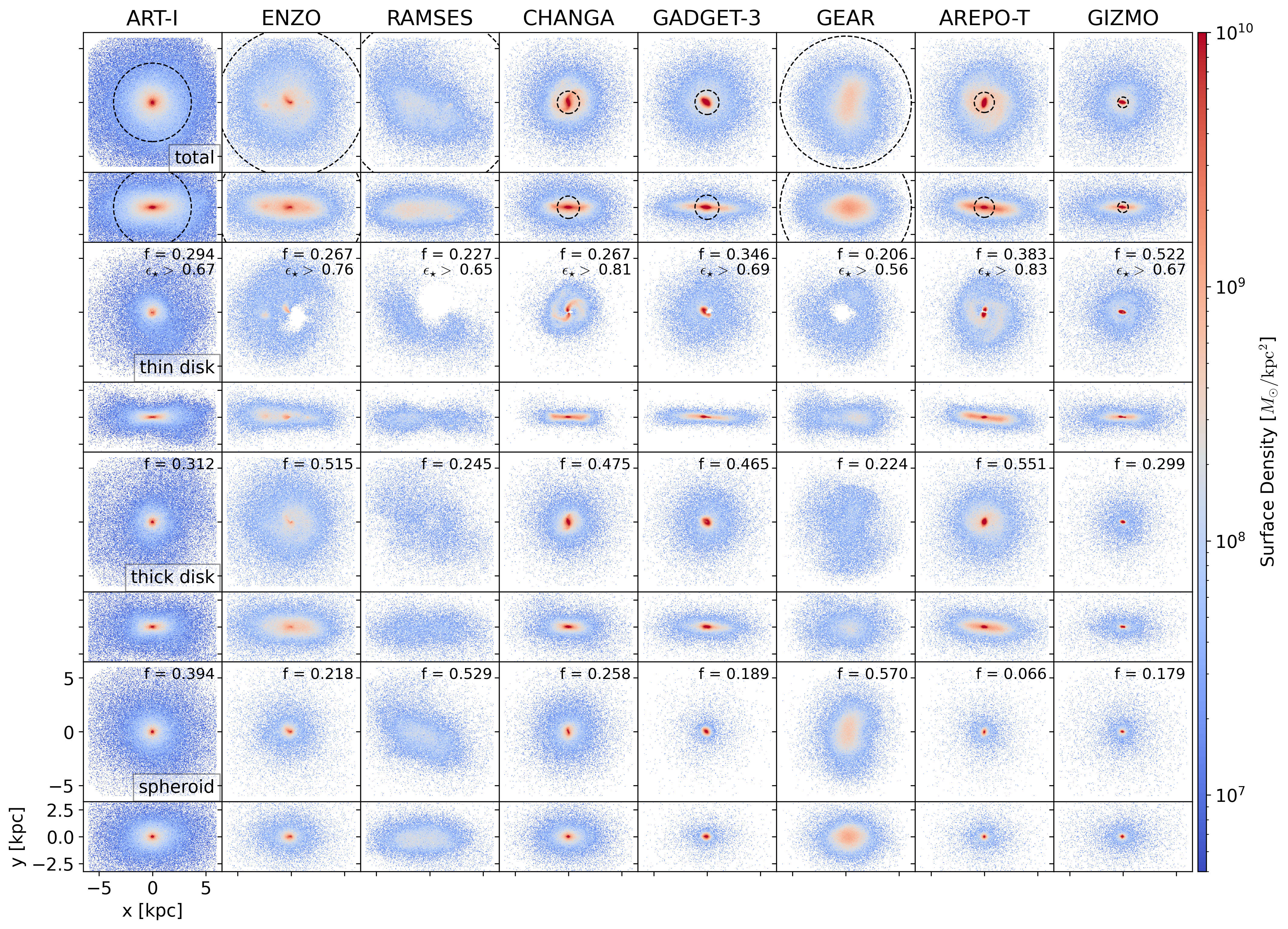}
    \caption{Face-on and edge-on views of stellar surface density at redshift $z=2.8$. 
    From {\it top} to {\it bottom}, each set of rows represents the total, thin disk, thick disk, and spheroid stellar component. 
    The mass fractions of each component are shown in the upper right corner of each panel, and the circularity threshold between the thin and thick disk, $\epsilon_{\text{thre}}$, is displayed below the mass fraction of the thin disk component. The black dashed circles in the top panels represent $3 \times r_{1/2}$. 
    The results at $z=1$ are presented in Appendix \ref{sec:z1}. 
    For more information, see Section \ref{sec:disk_id}.}
    \label{fig:disk_decomp_z28}
    \vspace{2mm}
\end{figure*}

\subsection{The AGORA CosmoRun Suite}\label{sec:suite}

The \textit{CosmoRun} is a suite of high-resolution cosmological zoom-in simulations of a Milky Way-mass halo ($10^{12}\msun$ at $z=0$) across multiple code platforms. All codes started from an identical initial condition at $z=100$ and reached $z\leq 2$. 
The eight participating codes include those with various hydrodynamic solvers classified as AMR (\textsc{Art-I}, \textsc{Enzo}, \textsc{Ramses}), SPH (\textsc{Changa}, \textsc{Gadget-3}, \textsc{Gear}), and hybrid methods (\textsc{Arepo-t}, \textsc{Gizmo}). Here we denote \textsc{Arepo-t} as a \textsc{Arepo} code with \textit{thermal} feedback prescription. The results with the feedback used in \textsc{IllustrisTng}, denoted as \textsc{Arepo-tng}, are shown in Appendix \ref{sec:tng}. The most notable code-independent factors are the choices of stellar feedback strength, feedback scheme, and hydrodynamic solver type (AMR, SPH, or hybrid). After calibration, all codes produced similar stellar mass and stellar properties at $z \geq 4$ (Paper III). Detailed descriptions of each participating code are provided in Papers III and IV. 

The mass resolution in the highest-resolution region of the initial condition is $m_{\text{DM, IC}} = 2.8 \times 10^5 \msun$ and $m_{\text{gas, IC}} = 5.65 \times 10^4 \msun$ for dark matter and gas, respectively. A gravitational softening length of 800 comoving pc at $z>9$ and 80 proper pc afterward is adopted in SPH codes and hybrid methods. In the case of AMR codes, the finest cell size is set to 163 comoving pc. For details on runtime parameters, see Paper III.

We identified a slight timing discrepancy in the growth of the target galaxy, where the same merger event occurs at slightly different cosmic times (see Paper IV). To address this, slightly different redshifts are adopted for each code to ensure that all simulations reflect the same stage in the galaxy's growth history (Papers IV and V). In this paper, we instead use $z=2.8$ and $z=1$ as key redshifts for the target galaxy.\footnote{To minimize the impact of a minor merger at $z=1$, $z=1.03$ is adopted for \textsc{Gear}.} The redshift $z=2.8$ represents the latest point at which all codes remain unaffected by the merger that occurred at $z=1.85-2.35$, depending on the codes (see Table 2 in Paper IV). Moreover, it is also far enough from the previous major merger at $z\sim4.5$, which is expected to have settled disk kinematics. Note that six out of eight participating codes reach redshifts as low as $z = 1$, while \textsc{Ramses} and \textsc{Gizmo} end their simulations at $z=2$. It should be noted that the lowest redshift achieved by each code group does not reflect the performance of the code, but rather the availability of manpower and CPU time at the computing facilities accessible to each code group.

As highlighted in Paper VI, we remind readers that each participating code may not represent the full diversity of its broader code community, as variations in stellar feedback schemes and user-defined parameters can exist within each code. Therefore, simulation groups using the \textit{AGORA} codes with alternative feedback implementations should exercise caution when comparing their simulations to the \textit{CosmoRun} results for that code.

\vspace{5mm}

\subsection{Disk Identification}\label{sec:disk_id}

Disk stars in the simulations are identified based on their kinematic properties. \cite{2003ApJ...597...21A} defined the circularity parameter as $\epsilon_{\star} = J_z/J(E)$, where $J(E)$ is the maximum angular momentum at a given specific binding energy. Stars with higher circularities are classified as disk stars, whereas the bulge component exhibits a symmetric distribution with zero net circularity \citep{2003ApJ...597...21A, 2014MNRAS.437.1750M}. This method was later improved by introducing an additional kinematic parameter, $\eta_{\star}=J_p / J(E)$, where $J_p$ is the residual component of the angular momentum. In this method, disk stars are identified in a multi-dimensional parameter space defined by binding energy, $\epsilon_{\star}$, and $\eta_{\star}$ \citep{2012MNRAS.421.2510D, 2024ApJS..271....1Y}. 

\begin{figure*}
    \centering
    \vspace{2mm}       
    \includegraphics[width = \linewidth]{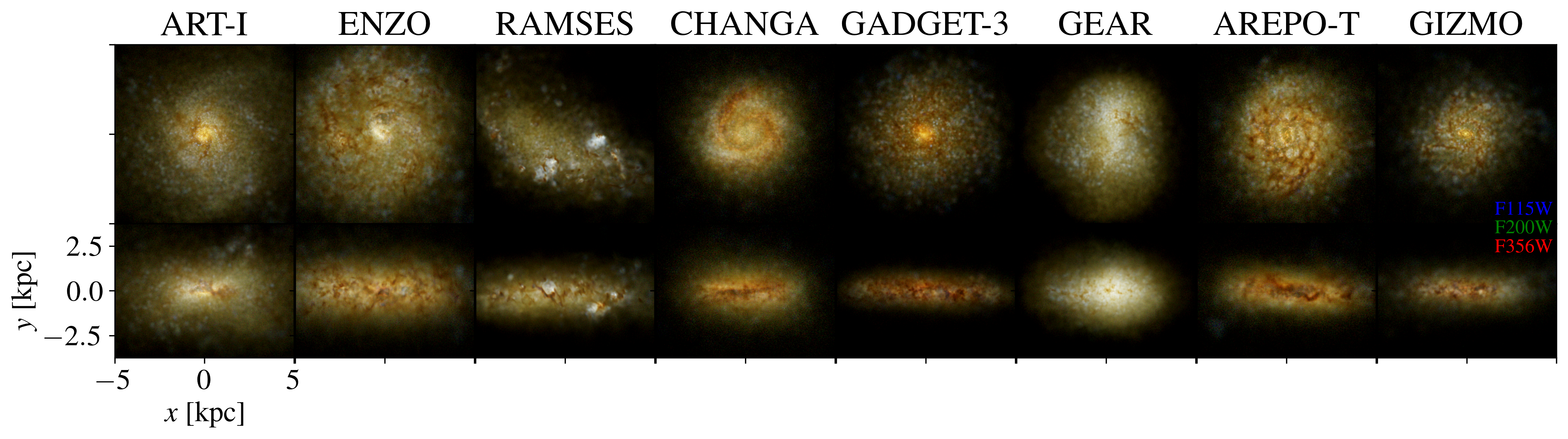}
    \caption{Mock JWST NIRCam RGB images (R: F356W, G: F200W, B: F115W) at $z=2.8$. 
    Upper panels and lower panels represent the face-on and edge-on views, the same viewing angles with Figure \ref{fig:disk_decomp_z28}.
    For more information, see Section \ref{sec:mock}.} 
    \label{fig:mock_z28} 
    \vspace{2mm}
\end{figure*}

Specifically, we follow \cite{2024arXiv240314749L} for the kinematic decomposition of the stellar component. In this method, circularity, binding energy, and $\eta_{\star}$ are used. The spheroidal component is defined as stars with negative $\epsilon_{\star}$ and their mirrored projection (negative circularity inverted) in three-dimensional parameter space, while the disk is defined by a positive excess of circularity. There are a few modifications from the original method due to technical difficulties. Firstly, the center of the galaxy, except in \textsc{Ramses} and \textsc{Gear}, is determined using the {\it shrinking sphere} method described by \cite{2003MNRAS.338...14P}. However, the centers in \textsc{Ramses} and \textsc{Gear} are less distinct due to their extended mass distributions, which impede the identification of a consistent center across multiple snapshots using the above method. Instead, we define the center as that of a spherical baryonic overdensity, which is 2000 times the critical density of the Universe, following the method described by \cite{Nguyen:inprep-a} (hereafter Paper IX). Visual inspection confirms that the two methods produce consistent centers for the remaining codes. Secondly, we assign stellar particles within $0.15 R_{\text{vir}}$ to the target galaxy, rather than assigning stars with the \texttt{SUBFIND} algorithm \citep{2001MNRAS.328..726S}. The definition of virial radius, $R_{\text{vir}}$, follows from \cite{1998ApJ...495...80B}. 

Unbound stars with $E>0$ or $\max(|\epsilon_{\star}|, |\eta_{\star}|)>1.5$ are excluded from the decomposition process \citep{2024arXiv240314749L}. The disk-to-total ratio (hereafter D/T) is defined as the mass fraction of stellar particles assigned to the disk, where the total stellar mass, $M_{\star}$, is defined as the mass of stellar particles within $0.15 R_{\text{vir}}$ but removing the contribution of the unbound stars. Finally, a two-component Gaussian mixture model is used in the parameter space to decompose disk stars into two groups. The threshold for classifying thin and thick disks, $\epsilon_{\text{thre}}$, is determined as the intersection point of the two Gaussian models along the $\eta_{\star}$ direction. All disk stars with $\epsilon_{\star} > \epsilon_{\text{thre}}$ are classified as thin disk, while the remainder are classified as thick disk. The spheroidal component could be further decomposed into bulge and halo \citep[e.g.,][]{2022MNRAS.515.1524Z,2024arXiv240314749L}. However, this decomposition is less straightforward for some codes at high redshift, and therefore we have chosen not to further decompose the spheroidal component. 

Readers should note that, rather than adopting a fixed $\epsilon_{\text{thre}}$ of around 0.7–0.8 \citep[e.g.,][]{2023MNRAS.523.6220Y, 10.1093/mnrasl/slad103}, the threshold varies among target galaxies, with $\epsilon_{\text{thre}} = 0.45$--$0.87$, after the end of the last major merger at $z\sim4.5$. \my{With this approach, we are able to identify a thinner component even at higher redshift, when discs are generally kinematically hotter (see \cite{2024arXiv240314749L} for further discussion).} In general, galaxies with lower $\epsilon_{\text{thre}}$ tend to have thin disks with larger scale heights than those with higher $\epsilon_{\text{thre}}$ \citep{2024arXiv240314749L}.\footnote{We note that, although the thin disk is defined as a kinematically colder component, its scale height is much larger than that of the MW, which is 300--400 pc.} 

An apparent limitation of this disk identification method is that during a major merger, stellar particles aligning with the orbital angular momentum may be misclassified as disk stars, potentially leading to an artificially high disk fraction. In this work, we distinguish the merger periods at $z \sim 4.5$ and $z \sim 2.2$; the D/T values during these periods---especially transient spikes---should be interpreted with extra caution.

\begin{figure*}
    \centering
    \vspace{2mm}
    \includegraphics[width = 0.86\linewidth]{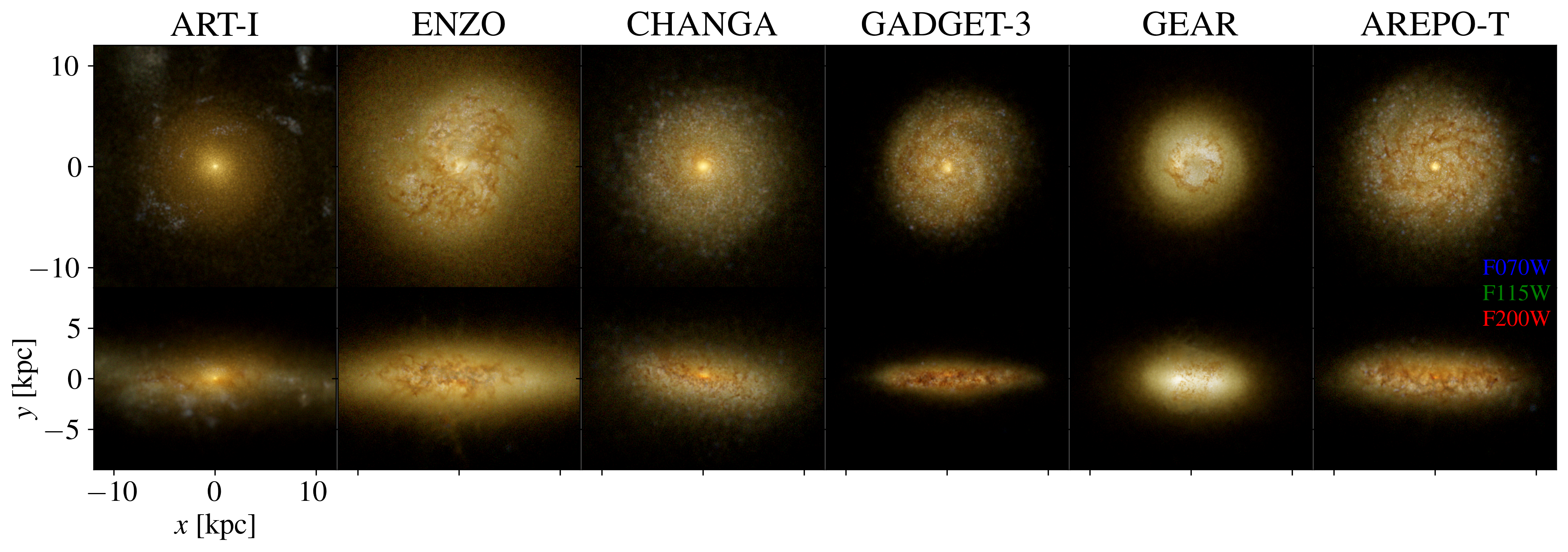}
    \vspace{-2mm}      
    \caption{Same as Figure \ref{fig:mock_z28} but with different broadband filters (R: F200W, G: F115W, B: F070W) at $z = 1$.
    Note that six out of eight codes seen in Figure \ref{fig:mock_z28} reach redshift $z = 1$, while \textsc{Ramses} and \textsc{Gizmo} end their simulations at $z=2$ (see Sections \ref{sec:suite} and \ref{sec:mock}). } 
    \label{fig:mock_z1}
    \vspace{4mm}
\end{figure*}

Figure \ref{fig:disk_decomp_z28} presents face-on and edge-on views of the stellar surface density distributions at $z=2.8$, along with the results of stellar kinematic decomposition. The face-on and edge-on views are based on the stellar angular momentum of all stars within a spherical radius of $r < 0.15 R_{\text{vir}}$. Three times the half-mass radius, $3 \times r_{1/2}$, is shown for comparison.\footnote{$r_{1/2}$ is defined as the radius that encloses half of the total stellar mass within $0.15 R_{\rm vir}$.} 
The mass fraction of the spheroidal component varies between codes from 0.066 to 0.570, indicating differences in disk fraction at $z=2.8$. In most codes, the thin disks are indeed represented with the morphology we typically envision. In some codes, however---such as \textsc{Ramses}, and \textsc{Gear}---the thin disks are confined to the outer regions, indicating that the inner stars fail to settle into circular orbits. The thin disk in \textsc{Enzo} exhibits azimuthal anisotropy, attributed to the ongoing formation of clumps. 

\subsection{Mock Observations}\label{sec:mock}

The mock images of the target galaxy at $z=2.8$ and $z=1$ (Figures 2 and 3) are generated with the radiative transfer code \texttt{SKIRT} \citep{2020A&C....3100381C}, for the analysis in Section \ref{sec:observation}. The gas and stellar components within a $0.15 R_{\rm vir}$ sphere are imported. Gas media from AMR and SPH codes are imported into \texttt{SKIRT} in their natural formats. Additionally, we treat the gas medium in \textsc{Gizmo} as SPH particles and in \textsc{Arepo-t} as Voronoi cells. An octtree grid is constructed to discretize the dust distribution. Dust production is assumed to be proportional to the gas mass if the temperature of a gas parcel is less than $T_{\rm max}$.

\begin{equation}\label{eqn_dust}
	\rho_{\rm dust} = 
	  \begin{cases}
    f_{\rm dust}Z_{\rm gas}\rho_{\rm gas} & \text{if $T < T_{\rm max}$} \\
    0 & \text{else}
    	\end{cases}
\end{equation}

Here $\rho_{\rm dust}$ and $\rho_{\rm gas}$ are the densities of dust and gas, respectively; $f_{\rm dust}$ is the dust-to-metal ratio; and $Z_{\rm gas}$ is the gas metallicity. In this study, we adopt $T_{\rm max} = 1.5\times 10^4 \, \mathrm{K}$ and $f_{\rm dust} = 0.2$. The choice of $T_{\rm max} = 1.5\times 10^4 \, \mathrm{K}$ ensures that the gas plateau at $\sim 10^4 \, \mathrm{K}$ in the density-temperature plane is properly captured (see Paper III). The selection of $f_{\rm dust} = 0.2$ follows \cite{2024A&A...683A.181B}.  

The Chabrier initial mass function \citep{2003PASP..115..763C} is used to estimate the stellar luminosity. The initial mass, metallicity, and age of the stellar particles are provided as inputs. The initial stellar mass is either obtained from the existing initial stellar mass field or estimated based on the stellar mass-loss implementation of each code. The distance to the 32nd nearest stellar particle is used as the smoothing length for softening, with the maximum smoothing length limited to $800 \pc$, following \cite{2024A&A...683A.181B}. 

We display the edge-on and face-on views of the mock snapshots in Figures \ref{fig:mock_z28} and \ref{fig:mock_z1}. RGB colors are generated using three JWST/NIRCam broadband filters: F115W, F200W, and F356W for $z=2.8$, and F070W, F115W, and F200W for $z=1$. To identify the size and shape of the galaxy, the F200W filter is used for snapshots at $z=2.8$, and the F115W filter is used for snapshots at $z=1$, as they approximate the $V$ band in the rest frame.  Note again that six codes reach redshift $z = 1$, while \textsc{Ramses} and \textsc{Gizmo} end their simulations at $z=2$ (see Section \ref{sec:suite}). The mock snapshots without gas medium (i.e., transparent) are displayed in Appendix \ref{sec:transparent}.

\subsection{Morphology Measurements}\label{sec:morphology}

We use the stellar half-mass radius, $r_{1/2}$, as the primary indicator of galaxy size. For this paper, we denote $r$ and $R$ as 3D radius and 2D radius in disk plane, respectively. We note, however, that the half-mass radius can be biased toward smaller values after gas compaction, especially in the presence of a dense core. While this metric is useful when tracing the formation of the inner core and its physical effects, it may underestimate the galaxy size by failing to encompass the star-forming gaseous disk. Measuring galaxy size based on the star-forming gas can provide an alternative solution \citep{2020MNRAS.493...87T, 2024A&A...682A.110B}. Therefore, we additionally define the size of a star-forming disk, $r_{\rm SFR}$, as twice the radius encompassing half of the total star formation rate, averaged over 100 Myr, following \cite{2024ApJ...972...73S}. Finally, the 2D projected half-light radius, $R_{e}$, is estimated from the face-on view of the mock snapshot and defined as the semi-major axis length of the ellipse that encloses half of the model light distribution. We use \texttt{galfit} \citep{2002AJ....124..266P} to fit a 2D Sérsic profile. 

The shape of a galaxy is characterized by $a$, $b$, and $c$ --- the lengths of the semi-axes of an ellipsoid, with $a \geq b \geq c$. We first determine the lengths of the semi-major and semi-minor axes, $A$ and $B$, from the 2D Sérsic profile, where $A$ is equivalent to $R_{e}$ and $B/A$ is assumed to correspond to $b/a$. Next, we generate an edge-on view, using the total stellar angular momentum and the semi-major axis as the projected plane.\footnote{These edge-on views are taken from different viewing angles than those in Figures \ref{fig:disk_decomp_z28}, \ref{fig:mock_z28}, and \ref{fig:mock_z1}.} From this projected snapshot, we fit another 2D Sérsic profile. The new ratio of the lengths of the semi-major and semi-minor axes, $B'/A'$, where $A'$ and $B'$ are the lengths of the semi-axes measured in the edge-on view, is regarded as $c/a$. We compare the sizes and shapes of the galaxies with observations in Section \ref{sec:observation}.

We use the concentration, $C_{82} = 5\log{(r_{80}/r_{20})}$, where $r_{80}$ and $r_{20}$ are the radii enclosing 80\% and 20\% of the total stellar mass, respectively, as an indicator of the stellar mass distribution.

\my{Finally, the $Q$ stability parameter is measured for the analysis in Sections \ref{sec:dis_compaction} and \ref{sec:dis_timediff}. We follow the method described in \cite{2023MNRAS.524.4346J}, which includes a disk thickness correction suggested by \cite{2013MNRAS.433.1389R}.}
\begin{equation}
T_i = \left\lbrace 
\begin{array}{@{}l@{\quad}l@{}}
1 + 0.6\left(\dfrac{\sigma_{i,z}}{\sigma_{i,r}}\right)^2 & \text{if } \left(\dfrac{\sigma_{i,z}}{\sigma_{i,r}}\right) < 0.5, \\
0.8 + 0.7\left(\dfrac{\sigma_{i,z}}{\sigma_{i,r}}\right) & \text{if } \left(\dfrac{\sigma_{i,z}}{\sigma_{i,r}}\right) > 0.5.
\end{array}
\right.
\end{equation}
\my{where $Q_i = T_i Q_{i,\mathrm{Toomre}}$ for $i = \mathrm{gas}$ and $\mathrm{star}$. Here, $Q_{i,\mathrm{Toomre}}$ denotes the standard Toomre $Q$ parameter for stars and gas \citep{1964ApJ...139.1217T}:}
\begin{equation}
Q_{\rm gas, Toomre} = \frac{\kappa \sigma_{r,\rm gas}}{\pi G \Sigma_{\rm gas}},\; Q_{\rm star, Toomre} = \frac{\kappa \sigma_{r,\rm star}}{3.36 G \Sigma_{\rm star}}
\end{equation}
\my{Here, $\kappa$ is the epicyclic frequency, $\sigma_r$ is the radial velocity dispersion, $G$ is the gravitational constant, and $\Sigma$ is the surface density of the disk. In the case of $Q_{\mathrm{star}, \mathrm{Toomre}}$, $\Sigma_{\mathrm{star}}$ excludes the contribution from the spheroidal stellar component. The two-component $Q$ stability parameter is adopted using the formulation of \cite{2011MNRAS.416.1191R}:}
\begin{equation}
\frac{1}{Q} = 
\left\lbrace 
\begin{array}{@{}l@{\quad}l@{}}
\frac{W}{Q_{\rm star}} + \frac{1}{Q_{\text{gas}}} & \text{if } Q_{\rm star} > Q_{\text{gas}}, \\
\frac{1}{Q_{\rm star}} + \frac{W}{Q_{\text{gas}}} & \text{if } Q_{\rm star} < Q_{\text{gas}}.
\end{array}
\right.
\end{equation}
\my{where}
\begin{equation}
W = \frac{2\sigma_{r, \rm gas}\sigma_{r, \rm star}}{\sigma_{r, \rm gas}^2 + \sigma_{r, \rm star}^2}
\end{equation}


\section{Results}\label{sec:result}

\subsection{Growth of Disk Over Cosmic Time}\label{sec:disk_growth}

\begin{figure*}
    \centering
    \includegraphics[width = 0.93\linewidth]{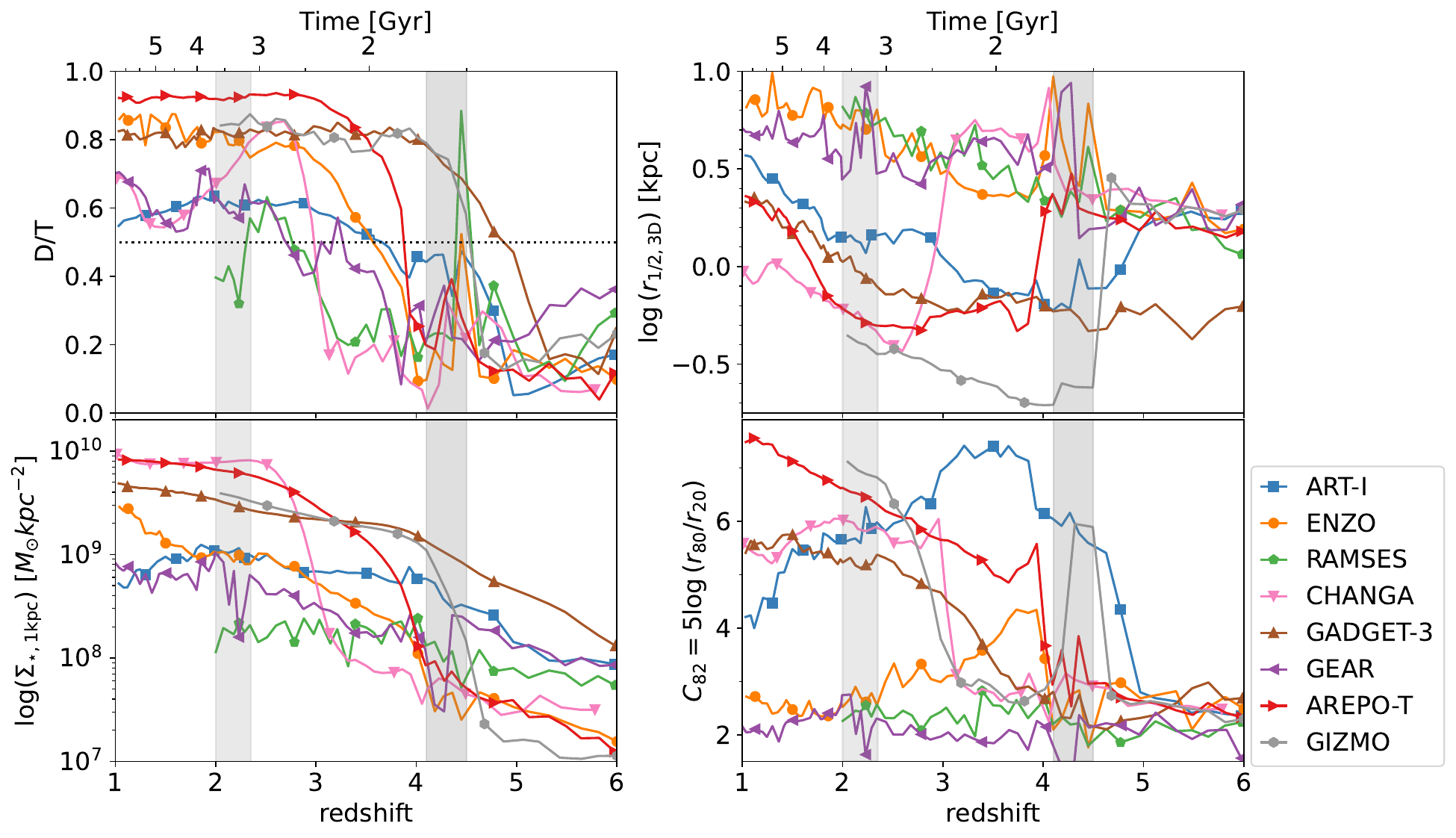}
    \caption{Evolution of disk properties of the target galaxy in each code. 
    The panels display the disk-to-total ratio (D/T), the half-mass radius ($r_{1/2}$), the stellar surface density within a $1 \kpc$ radius ($\Sigma_{{\star}, 1 \kpc}$), and the concentration ($C_{82}$; see Section \ref{sec:morphology}). 
    The gray-shaded regions indicate mergers at $z \sim 4.5$ and $z \sim 2.2$, during which D/T may not be well-defined. 
    The D/T of the target galaxies exceeds 0.5 before the merger event at $z \sim 2.2$, roughly coinciding with the increases in $\Sigma_{{\star}, 1 \kpc}$ and $C_{82}$, and the decrease in $r_{1/2}$.
    For more information, see Section \ref{sec:disk_growth}.}
    \label{fig:disk_evol}
    \vspace{3mm}
\end{figure*}

The top left panel in Figure \ref{fig:disk_evol} illustrates the evolution of the D/T over cosmic time. As depicted in the top left panel, the target galaxies achieve a D/T exceeding 0.5 before the merger events at $z \sim 2.2$ (indicated by a left gray-shaded region), indicating that the galaxies become rotation-dominated in their kinematic morphology. \my{This transition---formation of disk---occurs at different cosmic times, ranging from $z\sim 2.8$ (\textsc{Ramses} and \textsc{Gear}) to $z\sim 5$ (\textsc{Gadget-3}).} The spikes during the mergers, particularly at $z\sim 4.5$, are due to the insufficient measurement of D/T in strong turbulence and do not represent the physical fraction of the disk.\footnote{The halo mass ratios of the mergers at $z\sim 4.5$ and $z\sim 2.2$ are $\sim 0.6$ and $\sim 0.1$, respectively \citep{Nguyen:inprep-a}. Therefore, the first merger has a greater impact on the galaxy morphology.} After $z = 2.5$, D/T values remain steady across all codes except for \textsc{Changa} and \textsc{Ramses}, where they decline. Similar saturation is observed in the growth of stellar surface density within $1 \kpc$ (bottom left panel in Figure \ref{fig:disk_evol}), although \textsc{Enzo} continues to grow after $z = 1.5$. An oscillation in D/T is observed in \textsc{Gear} after $z=2$, which, based on visual inspection, is driven by minor mergers.

In Paper IV, we observed a rapid shift toward a more compact stellar distribution in \textsc{Changa}, \textsc{Arepo-t}, and \textsc{Gizmo}, coinciding with periods of star formation bursts (see Figure 7 in Paper IV). The rapid growth of D/T is simultaneously observed in these codes, with a formation timescale of less than 300 Myr (top left panel in Figure \ref{fig:disk_evol}). In these codes, the galaxy's half-mass radius decreases significantly to $r_{1/2} \lesssim 0.5 \kpc$, while stellar surface density within $1 \kpc$, $\Sigma_{{\star}, 1 \kpc}$, increases as stars form in the galaxy's inner region. \my{The increase in D/T indicates that the inner core formed during compaction is rotationally dominated.} This size contraction is also accompanied by an increase in the concentration parameter (right panel in Figure \ref{fig:disk_evol}). Following compaction, the concentration increases sharply to $C_{82} > 5$, indicating a more center-concentrated profile. \my{The increase in $C_{82}$ generally coincides with size compaction in most codes. In \textsc{Gizmo}, $C_{82}$ remains steady---excluding short-lived fluctuations during the merger---and increases about a Gyr after compaction. This is because both $r_{20}$ and $r_{80}$ drop sharply during compaction, whereas in other codes, $r_{80}$ remains steady, leading to an increase in $C_{82}$.} \del{While the increase in $C_{82}$ generally coincides with size compaction in most codes, in \textsc{Gizmo}, $C_{82}$ increases a Gyr after compaction, excluding short-lived fluctuations during the merger. The increase in D/T indicates that the inner core formed during compaction is rotationally dominated.}  

The rapid compactions and disk formation in the three codes --- \textsc{Changa}, \textsc{Arepo-t}, and \textsc{Gizmo} --- are apparently induced by the major merger at $z \sim 4.5$, represented by the thick gray-shaded region in Figure \ref{fig:disk_evol}. As mentioned in the Section \ref{sec:intro}, gas loses angular momentum and rapidly accretes into the galaxy during wet mergers, leading to starbursts in the galaxy's inner region---a process often referred to as \textit{wet compaction} \citep{2023MNRAS.522.4515L}. The significant time differences in \textsc{Changa}'s compaction following the merger suggest that \textsc{Changa} follows slightly different pathways through compaction. A detailed discussion of the compaction mechanism will be presented in Section \ref{sec:dis_timediff}. 

The other codes also exhibit signs of compaction, though to a lesser degree. In \textsc{Art-I}, the target galaxy undergoes compaction at $z \sim 4.8$, similar to the three previously mentioned codes, but with a less pronounced increase in D/T and stellar surface density. In \textsc{Gadget-3}, the galaxy reaches a D/T greater than 0.5 at the earliest epoch, with a notable size difference from the other codes already apparent at $z > 5$. However, the reduction in stellar size during inner core formation at $z \sim 5$ is less pronounced, and an increase in concentration appears at later time of $z \sim 3.5$, as $r_{80}$ expands due to star formation in the galaxy's outer regions. These transitions toward disk-like structures in \textsc{Art-I} and \textsc{Gadget-3} appear to be driven by mechanisms other than the major merger, such as frequent minor mergers in gas-rich environments.

\begin{figure*}[t]
    \centering
    \includegraphics[width=\linewidth]{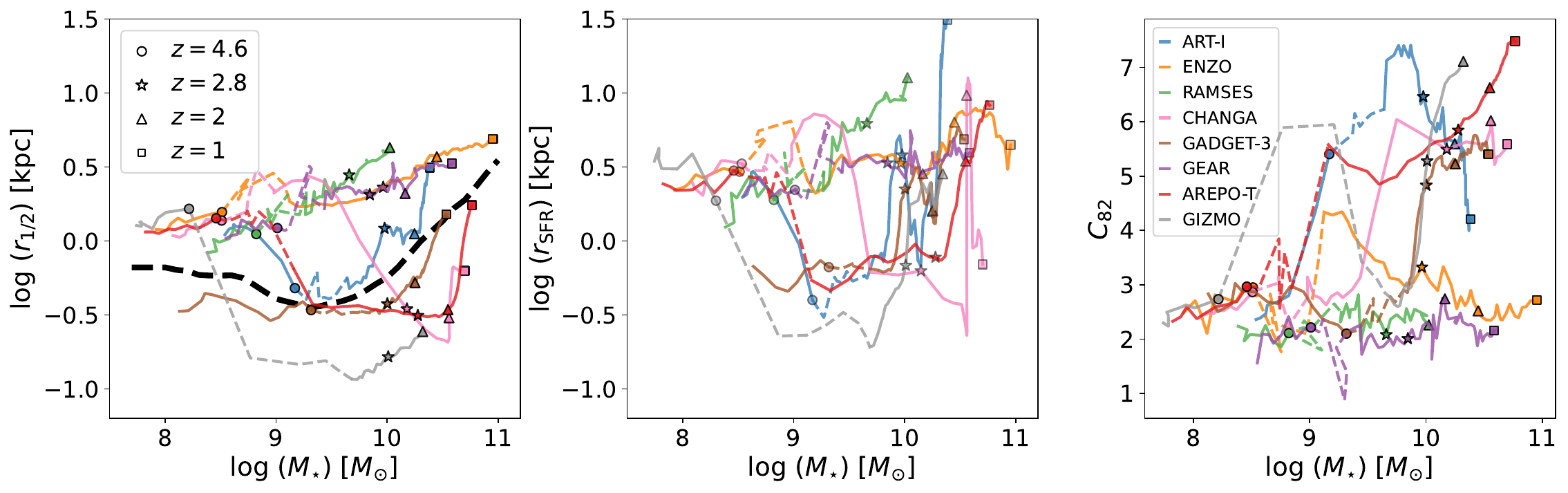}
    \caption{Evolution of the half-mass radius ($r_{1/2}$, {\it left panel}), the size of the star-forming disk ($r_{\rm SFR}$, {\it middle panel}), and the concentration ($C_{82}$, {\it right panel}) as functions of stellar mass of the target galaxy in each code. 
    The {\it thick black dashed line} in the {\it left panel} represents the evolution of galaxies undergoing wet compaction in the \textsc{Vela3} simulations \citep{2023MNRAS.522.4515L}. 
    In both $r_{1/2}$ and $r_{\rm SFR}$, \textsc{Enzo}, \textsc{Ramses}, and \textsc{Gear} exhibit a steady increase in size with stellar mass, while the other five codes show a contraction phase followed by an expansion phase. 
    Differences in $C_{82}$ between these two code groups become apparent at $M_{\star} \gtrsim 10^{9.5} \msun$. 
    The {\it dashed portion} in each line indicates the period that is heavily influenced by the major merger at $z \sim 4.5$ (denoted by a {\it circle marker}). 
    For more information, see Section \ref{sec:merger}.}
    \label{fig:Mstar-size}
    \vspace{3mm}
\end{figure*}

The signs of compaction are weakest in the three codes, \textsc{Enzo}, \textsc{Ramses}, and \textsc{Gear}. In \textsc{Enzo}, $\Sigma_{{\star}, 1 \kpc}$ and \my{$C_{82}$} increase by a factor of $\sim2$ at $z = 4$, but the concentration later decreases as $r_{20}$ expands. Growth in D/T is observed during this period, but it occurs without any contraction in $r_{1/2}$. In contrast, \textsc{Ramses} and \textsc{Gear} show lower D/T values, with little evidence of size contraction or an increase in concentration. Overall, while some codes exhibit signs of wet compaction around $z \sim 4$, others show minimal or no such evidence. We delve deeper into the examination of compaction in the target galaxies in the following section.


\subsection{Wet Gas Compaction}\label{sec:merger}

To investigate the signatures of wet compaction in greater details, Figure \ref{fig:Mstar-size} presents the evolution of $r_{1/2}$ (left), $r_{\rm SFR}$ (middle) and $C_{82}$ (right) as functions of stellar mass. 
Key redshifts highlighted are $z=4.6$ (just before the major merger), $z=2.8$ (just before the merger at $z\sim2.2$), $z=2$, and $z=1$. 
We compare the growth of the target galaxies with the median evolutionary trend observed in the \textsc{Vela3} cosmological galaxy simulations  \citep[thick black dashed line;][]{2023MNRAS.522.4515L}, which found $r_{1/2} \sim 0.5$ kpc at a characteristic stellar mass of $10^{9.5}\msun$ \citep{2016MNRAS.458..242T, 2023MNRAS.522.4515L}.

In the left panel of Figure \ref{fig:Mstar-size}, two distinct evolutionary trends are evident across the simulation codes: (1) a continuous increase in $r_{1/2}$ with stellar mass (\textsc{Enzo}, \textsc{Ramses}, \textsc{Gear}) and (2) an initial contraction in stellar size followed by an expansion phase (\textsc{Art-i}, \textsc{Changa}, \textsc{Gadget-3}, \textsc{Arepo-t}, and \textsc{Gizmo}). 
The latter trend resembles that observed in the \textsc{Vela3} simulations, although the contraction phase is less pronounced in \textsc{Gadget-3}. \textsc{Gizmo}, on the other hand, shows the most violent contraction phase—$r_{1/2}$ decreases and $M_{\star}$ increases by $\sim 1.5$ dex during the major merger (gray dashed line in the left panel).
A nearly identical trend is observed in the evolution of $r_{\rm SFR}$ for stellar mass $M_{\star} < 10^{10}\msun$, although the decline in $r_{\rm SFR}$ is steeper.\footnote{When galaxies exceed this stellar mass, $r_{\rm SFR}$ fluctuates by $\sim$1 dex or more in \textsc{Art-i} and \textsc{Changa}, driven by periods of quenched star formation.} Since $r_{\rm SFR}$ serves as a proxy for the size of the star-forming disk, this suggests that the gaseous disk contracts with more shorter timescale during the compaction phase.
The two diverging code groups can also be seen in $C_{82}$ at $M_{\star} \gtrsim 10^{9.5} \msun$. 

As presented in the previous section, the compaction in the \textit{CosmoRun} marks a clear transition from dispersion-dominated to rotation-dominated systems, as demonstrated by e.g., \cite{2023MNRAS.522.4515L}. 
The two codes with minimal signs of compaction, \textsc{Ramses} and \textsc{Gear}, exhibit notably lower D/T values, not exceeding 0.5 until $z=3$ (Figure \ref{fig:disk_evol}). Interestingly, \textsc{Enzo} forms a rotationally supported disk without a strong signature of compaction. Disk formation in \textsc{Enzo} differs from that in other codes, as the inner mass concentration is obscured in the half-mass radius metric due to clump formation within the disk plane (Figure \ref{fig:disk_decomp_z28}). Furthermore, the increase in $C_{82}$ during the merger is mitigated by continued star formation in the outer disk regions after the merger. 

\begin{figure*}[t]
    \centering
    \includegraphics[width=\linewidth]{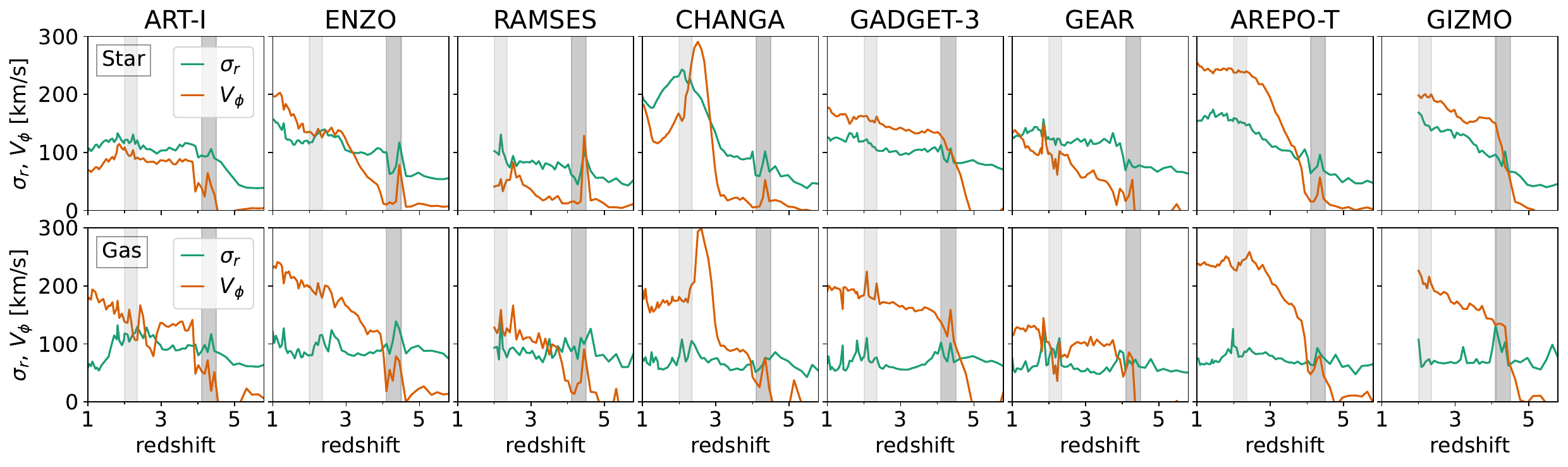}
    \caption{\my{Radial velocity dispersion ($\sigma_{r}$, \textit{green}) and tangential velocity ($V_{\phi}$, \textit{orange}) of the stellar and gas components as a function of cosmic time. The gray-shaded regions indicate mergers at $z \sim 4.5$ and $z \sim 2.2$. A rapid transition from $V_{\phi}/\sigma_{r} < 1$ to $V_{\phi}/\sigma_{r} \gtrsim 2$ is highlighted for some codes, primarily induced by the major merger at $z \sim 4.5$. For more details, see Section \ref{sec:merger}.} } 
    \label{fig:vcirc-sigma}
    \vspace{3mm}
\end{figure*}

\my{As a secondary metric of compaction, we examine the radial velocity dispersion ($\sigma_r$) and tangential velocity ($V_\phi$) over cosmic time (Figure \ref{fig:vcirc-sigma}). Both $V_\phi$ and $\sigma_r$ are calculated as the mass-weighted mean velocity and dispersion in cylindrical coordinates aligned with the stellar angular momentum. The evolution of stellar $V_\phi$ closely matches that of the D/T ratio in Figure \ref{fig:disk_evol}, reflecting the fact that the disk component is kinematically defined as rotationally dominated stars. For gas, $\sigma_r$ remains generally steady across cosmic time in all codes, while stellar $\sigma_r$ either stays constant or increases over time in \textsc{Changa}, \textsc{Arepo-t}, and \textsc{Gizmo}. A rapid transition in the gas component from $V_\phi/\sigma_r < 1$ to $V_\phi/\sigma_r \gtrsim 2$ occurs after the major merger at $z \sim 4.5$. Exceptions are \textsc{Art-I}, \textsc{Ramses}, and \textsc{Gear}, where the transition occurs but only to a relatively low $V_\phi/\sigma_r \sim 1$–2 for $2 < z < 4$.
}

\my{These transitions in kinematic properties are less highlighted in stellar component, especially in \textsc{Ramses} and \textsc{Gear}. \textsc{Enzo}, likewise, reach stellar $V_{\phi}/\sigma_{r} \sim 1$ at $z=3$, showing significant time delay after that in gas. We speculate that while these codes form rotating gas disk after the coherent gas inflow during the merger, it did not turn to the gas compaction and starburst. \textsc{Art-I}, finally, show rapid increase in $V_{\phi}$ at the major merger at $z=4.5$ but do not reach $V_{\phi}/\sigma_{r}$ until $z=1$, in line with low D/T shown in Figure \ref{fig:disk_evol}. These results support that the merger is responsible of formation of rotationally dominated gas disk, whereas the coherent formation of rotation-dominated stellar disks on short timescales is observed only in codes that exhibit strong signatures of compaction (\textsc{Art-I}, \textsc{Changa}, \textsc{Gadget-3}, \textsc{Arepo-T}, and \textsc{Gizmo}). The rapid rise of $V_{\phi}/\sigma_{r}$ in both stellar and gas components is qualitatively consistent with earlier studies of wet compaction \citep{2015MNRAS.450.2327Z, 2023MNRAS.522.4515L}. This support that the compaction in stellar size is accompanied by the kinetic transition, as discussed in the earlier section.} 



\my{Overall, these results support the idea that mergers are responsible for the formation of rotationally dominated gas disks, while the rapid and coherent formation of rotation-dominated stellar disks is observed only in codes with a strong signature of compaction (\textsc{Art-I}, \textsc{Changa}, \textsc{Gadget-3}, \textsc{Arepo-t}, and \textsc{Gizmo}). This further suggests that compaction in stellar size is accompanied by a kinematic transition, as discussed earlier.}

\my{In summary, wet compaction appears to act as a key transition point to rotationally dominated kinematics in some codes, though other evolutionary tracks without clear compaction also exist.} In the following section, we examine how the disk grows subsequently after compaction, contrasting codes that show clear signs of compaction (\textsc{Art-i}, \textsc{Changa}, \textsc{Gadget-3}, \textsc{Arepo-t}, and \textsc{Gizmo}) with those that show little to no compaction (\textsc{Enzo}, \textsc{Ramses}, and \textsc{Gear}). \my{This classification is based primarily on the evolution of stellar size; however, the two groups also differ in the growth of the star-forming disk and in the time evolution of $V_{\phi}/\sigma_{r}$. Variations within each group remain significant: for example, \textsc{Art-I} maintains lower D/T and $V_{\phi}/\sigma_{r}$ after compaction compared to other compaction-group codes; \textsc{Changa} shows a substantial time delay between the merger and the onset of compaction; and \textsc{Gadget-3} lacks a clear signature of stellar contraction despite a marked transition in $V_{\phi}/\sigma_{r}$. Despite these caveats, this grouping provides a useful framework for linking the disk formation channel to the resulting disk properties.}


\subsection{Stellar Age Gradients} \label{sec:age}

\begin{figure*}[t]
    \centering
    \includegraphics[width = 1.02\linewidth]{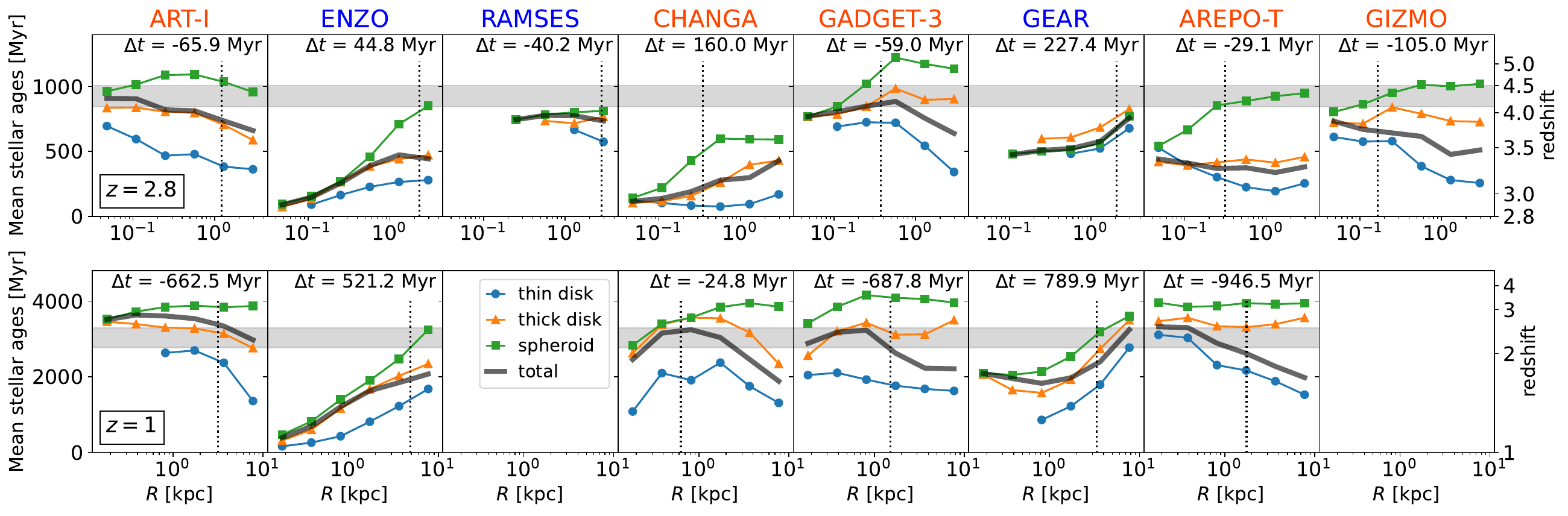}
    \caption{Radial stellar age profiles at $z=2.8$ ({\it top panels}) and $z=1$ ({\it bottom panels}) for the three kinematic stellar components --- thin disk, thick disk, and spheroid --- demonstrating the stellar age gradients.
    A {\it vertical dotted line} in each panel represent $r_{1/2}$. 
    A {\it gray shaded region} represents the mergers at $z\sim4.5$ ({\it top panels}) and $z\sim2.2$ ({\it bottom panels}). 
    The mean stellar age difference between the stars outside and inside $r_{1/2}$ is shown as $\Delta t$ in each panel. 
    The names of codes with strong signatures of compaction are labeled in orange, while those with weaker signatures are in blue. Only radial bins containing more than 30 stellar particles are displayed.
    Thin disk stars show negative stellar age gradients after compaction in codes with compact stellar sizes (\textsc{Art-I}, \textsc{Gadget-3}, \textsc{Arepo-t}, and \textsc{Gizmo}). 
    Note that six out of eight codes (seen in the upper panels) reach redshift $z = 1$ (bottom panels), while \textsc{Ramses} and \textsc{Gizmo} end their simulations at $z=2$ (see Sections \ref{sec:suite} and \ref{sec:mock}). 
    For more information, see Section \ref{sec:age}.} 
    \label{fig:age_histogram}
    \vspace{3mm}
\end{figure*}

By examining the spatial distribution of young and old stars in the disk components, we aim to understand how disk formation differs between the two code groups. The upper panels of Figure \ref{fig:age_histogram} illustrate the mean stellar age within radial annuli at $z=2.8$. Across all participating codes, the spheroidal components (green squares) exhibit either positive or flat age gradients (i.e., the mean stellar age increases or plateaus with distance from the center). In contrast, the thin and thick disk components (blue circles and orange triangles) display different behaviors depending on whether the codes exhibit strong signatures of compaction. The codes with weaker compaction signatures show positive or flat age gradients in the disk structures (codes labeled in blue in Figure \ref{fig:age_histogram}). Conversely, codes with apparent compaction generally exhibit negative age gradients in the thin disk and flat gradients in the thick disk, resulting in a negative net age gradient (black solid lines). The only exception is \textsc{Changa}, which exhibits a positive gradient despite forming a compact disk. Since in \textsc{Changa} the compaction occurred only a few hundred Myr before the target redshift of $z=2.8$, the inside-out growth that typically follows compaction is not captured at this time step. This is consistent with the positive or flat color gradients observed in post-starburst galaxies \citep[][]{2020MNRAS.497..389D,2021ApJ...915...87S}.

The differences in gradients associated with morphological shapes align with previous studies, which found that galaxies with spheroidal morphologies tend to have positive age gradients, while disk galaxies show negative gradients \citep{2022MNRAS.511.1072P}. However, it is worth noting that the distinction between the galaxies in different morphologies is not observed at $z=2$ in the earlier study, with mildly positive gradient with $\Delta t \sim 100$ Myr, where $\Delta t$ is the mean stellar age difference between the stars outside and inside of half-mass radius \citep[see Figure 5 in ][]{2022MNRAS.511.1072P}. The \textit{CosmoRun} galaxies, in contrast, show a clear separation in $\Delta t$ at $z=2.8$ between the two groups, except for \textsc{Changa}, as most \textit{AGORA} codes exhibit disk formation and spin-up at earlier times, around $z \sim 4$.

As shown in the lower panels of Figure \ref{fig:age_histogram}, similar trends persist at $z=1$ for the codes that reached this redshift. \textsc{Enzo} and \textsc{Gear} exhibit positive age gradients, while the other codes display flat or negative gradients. Interestingly, the galaxy in \textsc{Changa} exhibits a complex pattern, with a positive gradient within $R < r_{1/2}$ and a negative gradient beyond $r_{1/2}$. This behavior reflects active star formation in the central region of the galaxy near $z=1$. 

\begin{figure*}[t]
    \centering
    \includegraphics[width = \linewidth]{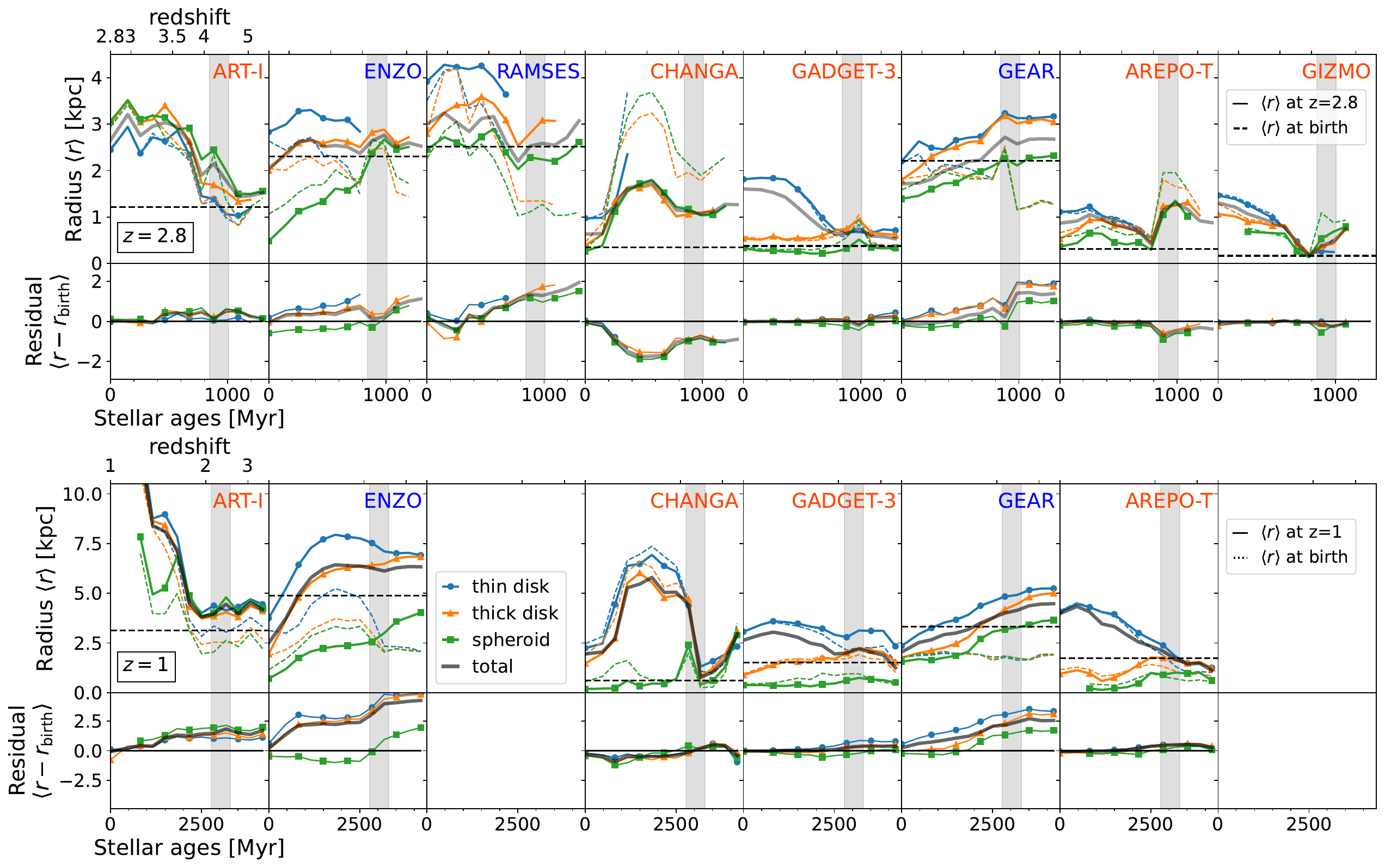}
    \caption{The {\it upper panels} show the mean radial positions of stellar particles from the galaxy center at $z=2.8$ ({\it solid lines}) and at their time of formation ({\it dashed lines}). 
    The three kinematic components --- thin disk, thick disk, and spheroid --- are displayed separately with the same color and shape convention as Figure \ref{fig:age_histogram}. 
    A {\it horizontal dashed line} in each panel represent $r_{1/2}$. 
    The {\it marginal plots} at the bottom show the residuals between the stellar radii at $z=2.8$ and their formation radii. 
    Only stars formed in situ are considered. 
    The lower panels present the same analysis at $z=1$.      
    A {\it gray shaded region} represent the mergers at $z\sim4.5$ ({\it top panels}) and $z\sim2.2$ ({\it bottom panels}).     
    The names of codes with strong signatures of compaction are labeled in orange, while those with weaker signatures are in blue. 
    A positive residual ({\it marginal plots}) indicates outward movement of stars after their formation. 
    Stars in the thin disk grow in an inside-out manner following compaction. 
    Stars formed at earlier epochs tend to migrate outward in codes with larger stellar sizes (\textsc{Enzo}, \textsc{Ramses}, \textsc{Gear}). 
    Note that six out of eight codes (seen in the upper panels) reach redshift $z = 1$ (bottom panels), while \textsc{Ramses} and \textsc{Gizmo} end their simulations at $z=2$ (see Sections \ref{sec:suite} and \ref{sec:mock}). 
    For more information, see Section \ref{sec:sfh}.} 
    \label{fig:mean_radius}
    \vspace{3mm}
\end{figure*}

Readers should note that we represent both in-situ and ex-situ stars in this analysis.\footnote{In-situ stars are defined as those formed within a $0.15 R_{\rm vir}$ sphere; ex-situ stars are all others.} We verify that the results with only in-situ stars are nearly identical, as the target galaxies are dominated by in-situ formation, comprising more than 90\% of the stellar mass. The kinematic properties of ex-situ stars and their contribution to the morphology will be explored in Paper IX. Finally, we caution that an age gradient does not imply a corresponding gas-phase metallicity gradient. Strong feedback can redistribute gas, moving metals from the inner to outer regions and thus flattening metallicity gradients \citep{2013A&A...554A..47G}. In Paper IV, we find that \textsc{Art-I}, \textsc{Ramses}, and \textsc{Changa} exhibit nearly flat radial metallicity profiles outside the very central region, whereas the other codes show negative gradients (see Figure 11 in Paper IV). Both \textsc{Art-I} and \textsc{Changa} have higher inner metallicities due to compaction but retain flat gradients in outer regions. In \textsc{Enzo} and \textsc{Gear}, positive stellar age gradients are associated with negative metallicity gradients, while in \textsc{Ramses}, a flat metallicity profile is observed. 

\subsection{Where In the Disk Did Stars Form?}\label{sec:sfh}

To investigate the origin of disparities in the age gradient, Figure \ref{fig:mean_radius} shows the mean distance of in-situ stars from the center at $z=2.8$ and $z=1$, binned by stellar age. The dashed lines represent the mean distance from the center at the time of formation for each stellar particle. By examining the mean residual between these two distances, a positive (negative) residual indicates that the stars have moved outward (inward) after their formation. 

\subsubsection{Compaction Followed by Inside-out Quenching}
Stars formed in the inner region, on the scale of $r_{1/2}$, during the wet compaction phase for the codes with strong signature of compaction (\textsc{Art-I}, \textsc{Gadget-3}, \textsc{Arepo-t}, and \textsc{Gizmo}). Sets of stellar particles formed after the compaction exhibit a higher mean radius, which increases as stellar age decreases (black solid lines after $z\sim4$ for the aforementioned codes). This trend is most prominent in the thin disk, while after compaction, the mean radii of the thick disk and spheroidal components show less significant evolution. Most codes with a strong signature of compaction exhibit stable orbits with minimal residuals between the distance at the time of star-formation and at $z=2.8$, whereas some codes, \textsc{Changa} and \textsc{Arepo-t}, display negative residuals prior to the compaction phase, reflecting strong inward accretion of star-forming gas, with stars inheriting the infalling kinematics of the gas. 

The rejuvenation events in \textsc{Changa} at $z=2.8$ and $z=1.2$ result in the complex gradient observed in Figure \ref{fig:age_histogram}, represented by a rapid decrease in the mean radial position of stellar particles. Variability in the star formation rate reflects the baryon cycle over a long timescale. The star formation rate in \textsc{Changa} peaked at $z=2.8$, during the compaction phase, shortly after which stellar feedback destroyed the gaseous disk and suppressed star formation, reaching its lowest value at the merger at $z\sim2.2$ (see Figure 7 in Paper IV). The gas expelled into the circumgalactic region is later reaccreted onto the galaxy, triggering another starburst at $z=1.2$. The evolution of the target galaxy in \textsc{Changa} resembles a \textit{`breathing'} process, although the interval between starbursts is several times longer than reported in previous studies of the FIRE suite \citep{2016ApJ...820..131E, 2017MNRAS.466...88S, 2024MNRAS.527.7871C}. The unique characteristics \textsc{Changa} exhibit will be further discussed in Section \ref{sec:dis_timediff}.

\subsubsection{Impact of Stellar Migration on Age Gradients}

\begin{figure}
    \centering
    \includegraphics[width=0.98\linewidth]{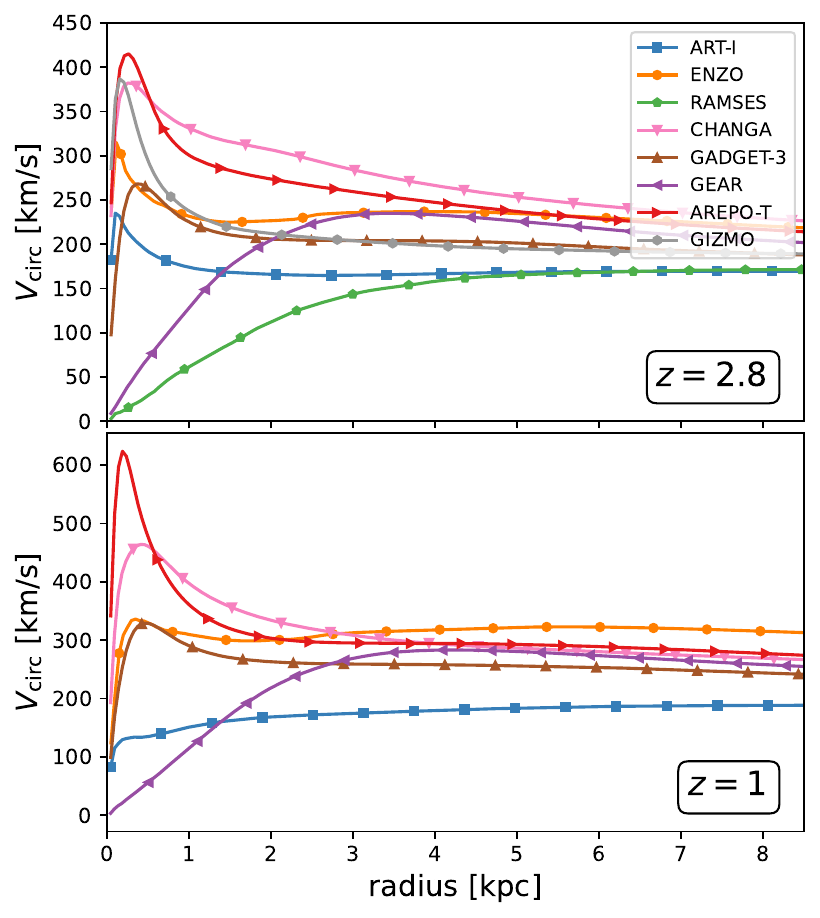}
    \caption{Circular velocity profiles, $\sqrt{GM(<r)/r}$, at $z=2.8$ ({\it top}) and $z=1$ ({\it bottom}), where $M(r)$ is the enclosed mass.
    Slow-rising rotation curves are observed in \textsc{Gear} and \textsc{Ramses}, while the other codes tend to exhibit steeply rising curves near the center.
    Note that six out of eight codes (seen in the upper panels) reach redshift $z = 1$ (bottom panels), while \textsc{Ramses} and \textsc{Gizmo} end their simulations at $z=2$ (see Sections \ref{sec:suite} and \ref{sec:mock}). 
    For more information, see Section \ref{sec:sfh}.}
    \label{fig:v_circ}
\end{figure}

When measured at the onset of star formation (dashed lines in Figure \ref{fig:mean_radius}), the three codes with weak signatures of compaction (\textsc{Enzo}, \textsc{Ramses}, \textsc{Gear}) display a similar trend: the distance from the center slightly increases over time. However, these stars display outward migration by $z=2.8$ (solid lines in the top panels and positive residuals in the marginal plots). This outward migration is the primary reason for the positive age gradients in these galaxies; neither \textsc{Ramses} nor \textsc{Gear} show signs of outside-in quenching, as indicated by the mean radial positions of stellar particles at their time of formation. Interestingly, \textsc{Enzo} deviates at lower redshift ($z<1.6$), where it undergoes outside-in quenching, further amplifying the positive stellar age gradient. 

The outside-in stellar growth in \textsc{Enzo} is also indicated by a decrease in $r_{\text{SFR}}$ from $8 \kpc$ to $4 \kpc$ (Figure \ref{fig:Mstar-size}) and by visual inspection, which shows that the star-forming, cold gas disk shrinks during this period. \textsc{Art-I} also displays mild outward migration, though strong inside-out quenching counteracts its effect on the overall gradient and results in negative age gradient. Mergers and stellar feedback could play a significant role, inducing fluctuations in the potential well and dynamical heating of the stars \citep[see][]{2016ApJ...820..131E,2019MNRAS.490.1186G}. Figure \ref{fig:v_circ} presents the circular velocity profiles, indicating that the gravitational potential wells in \textsc{Art-I}, \textsc{Ramses}, and \textsc{Gear} are shallower compared to those in the other codes.\footnote{The differences in circular velocities in \textsc{Art-I} and \textsc{Gear} at $z=1$ result from the lack of baryonic content in the galactic region, rather than from differences in the dark matter distribution induced by the timing discrepancy discussed in Papers IV and V. In other words, while the dark matter mass shows good convergence between the codes, the discrepancy arises due to baryonic effects.} This finding supports the idea that outward migration, driven by mergers and stellar feedback-induced dynamical heating, is mitigated when a strong central gravitational potential is present. \textsc{Enzo}, however, shows a rotation curve similar to those with negative age gradients. \textsc{Enzo} exhibits multiple massive clumps within the stellar disk, which undergo inward migration due to dynamical friction (see also the discussion in Section \ref{sec:dis_compaction} on \my{stability} analysis). On the other hand, the inner core is significantly redistributed and contributes to net outward migration. This differs from the expected evolution of clumps under dynamical friction, which typically leads to central concentration and bulge formation \citep{2013MNRAS.436..259P, 2024MNRAS.529.2702H}. Dynamical heating from other sources, such as mergers in a cosmological context, could explain these differences.


Finally, we note a correlation between disk fraction (D/T) and inner mass concentration, characterized by maximum circular velocity. \textsc{Art-I}, \textsc{Ramses}, and \textsc{Gear} exhibit both lower D/T and lower maximum circular velocity values, consistent with earlier findings by \cite{2023MNRAS.525.2241H}. The circularity parameter, $\epsilon_{\star}$, approaches unity for purely circular orbits at a given binding energy, resulting in a higher D/T ratio and minimal stellar migration in radial direction, which is observed in \textsc{Changa}, \textsc{Gadget-3}, \textsc{Arepo-t}, and \textsc{Gizmo}. 

\subsection{Comparison With Observations}\label{sec:observation}

\begin{table*}[t]
\centering
\vspace{1mm}
\caption{List of galaxy properties in the \textit{AGORA} \textit{CosmoRun} simulation suite at $z=2.8$ and $z=1$. }
\vspace{-2mm}
\renewcommand{\arraystretch}{1.2} 
\setlength{\tabcolsep}{8pt} 
\begin{tabular}{clcccccccc}
\hline
 redshift $z$   &                  & {\textsc{Art-I}} & {\textsc{Enzo}} & {\textsc{Ramses}} & {\textsc{Changa}} & {\textsc{Gadget-3}} & {\textsc{Gear}} & {\textsc{Arepo-t}} & {\textsc{Gizmo}} \\
\hline
\hline
$\mathbf{z=2.8}$ & $r_{1/2}$ [kpc]\tablenotemark{\scriptsize \textdagger}   & 1.22 & 2.30 & 2.79 & 0.35 & 0.38 & 2.05 & 0.31 & 0.16 \\
    (F200W)  & $r_{\rm SFR}$ [kpc] &  3.81  &   3.35     &   6.22    &   0.63     &   2.25       &   3.37   &    0.78      &    0.68      \\
          & $R_{e}$ [kpc]    & 2.51 & 2.71 & 2.65 & 1.84 & 2.30 & 1.80 & 2.16 & 2.06 \\
          & $n$  & 0.9 & 0.6 & 0.8 & 0.7 & 0.6 & 0.6 & 0.7 & 1.1  \\
          & $M_{\star}$ [$10^{10}\msun$]&    0.95     &   0.92        &    0.46    &  1.51      &   1.01   &  0.70    &   1.89     &   1.02     \\
\hline
$\mathbf{z=1}$   & $r_{1/2}$ [kpc]  &  3.14      &    4.88    &    -     &     0.63       &  1.52       &   3.34      &   1.75        &    -    \\
(F115W)    & $r_{\rm SFR}$ [kpc] &  \,\,-\tablenotemark{\scriptsize \textdaggerdbl}  &   3.22     &   -   &   0.69     &   3.94       &   3.95   &    7.63      &    -      \\
    & $R_{e}$ [kpc] &  3.85 & 5.91 & - & 4.00 & 3.43 & 2.70 & 4.30 &    -        \\
          & $n$    & 2.5 & 0.7 & - & 1.1 & 0.6 & 0.6 & 0.5 & -  \\
          & $M_{\star}$ [$10^{10}\msun$]&    2.42     &   8.98        &    -   &  4.99    &   3.43   & 3.85  &   5.83   & -   \\
\hline
\end{tabular}
\tablenotetext{\scriptsize $\textdagger$}{\scriptsize Each row represents the 3D stellar half-mass radius ($r_{1/2}$), the size of a star-forming disk ($r_{\rm SFR}$), the effective radius in face-on views with dust attenuation ($R_e$), Sérsic index ($n$), and the galaxy stellar mass ($M_{\star}$). }
\vspace{-2mm}
\tablenotetext{\scriptsize $\textdaggerdbl$}{\scriptsize Note that $r_{\rm SFR}$ of \textsc{Art-I} at $z=1$ is not represented due to the lack of star-forming gas inside the target galaxy. }
\label{tab:radii}
\vspace{1mm}
\end{table*}

\begin{figure*}
    \centering
    \includegraphics[width=\linewidth]{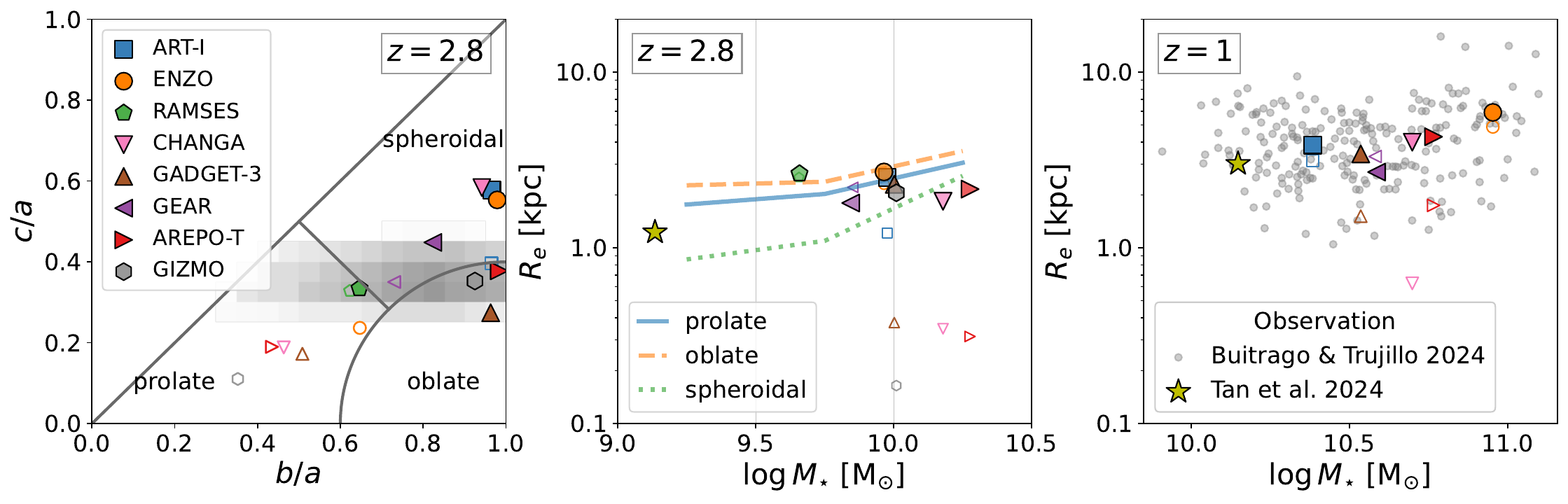}
    \caption{\textit{Left:} 3D axis ratios ($b/a$, $c/a$) of the target galaxy at $z=2.8$ with dust extinction effects ({\it filled markers}). 
    We also show the ratios based on stellar distribution without extinction ({\it small, open markers}). 
    A {\it gray-shaded region} indicates the observational results in the same redshift bin at $\log{(M_{\star}/\mathrm{M_{\odot}})} = 10.0$--$10.5$ \citep{2024ApJ...963...54P}.
    \textit{Middle:} Size--mass relation at $z=2.8$ using the effective radius with dust extinction effects, $R_e$, and the 3D stellar half-mass radius, $r_{1/2}$ ({\it filled} and {\it open markers}, respectively). 
    {\it Solid} and {\it dotted lines} represent the mean 3D size--mass relation at $2.5 < z < 3$ \citep{2024ApJ...963...54P}. 
    \textit{Right:} Size--mass relation at $z=1$, with the observational data in \cite{2024A&A...682A.110B}.   
    The yellow star in the middle and right panels represents the median $M_{\star}$ and $R_e$ of MW-like progenitors at each redshift \citep{2024arXiv241207829T}.
    For more information, see Section \ref{sec:observation}.}
    \label{fig:shape}
    \vspace{3mm}
\end{figure*}

During compaction, the sizes of galaxies contract, reaching $\log{r_{1/2}} \lesssim -0.5$, as a result of core formation in their inner regions (see Sections \ref{sec:disk_growth} and \ref{sec:merger}). While the formation of the inner core is effectively captured by the metric $r_{1/2}$, it underestimates the galaxy size due to the presence of an inner core in galaxies with strong signatures of compaction. Table \ref{tab:radii} lists the size of the galaxy with different definitions. Another size metric, $r_{\text{SFR}}$, is a factor of $\gtrsim 2$ larger than $r_{1/2}$ at $z=2.8$, except \textsc{Enzo} and \textsc{Gear}. This indicates that the star-forming gaseous disk extends beyond the scale of $r_{1/2}$. The $R_e$ of the galaxy --- the projected effective radius with the inclusion of dust attenuation --- is greatly larger than $r_{1/2}$, due to the inclusion of dust attenuation in gas-rich systems at $z=2.8$.\footnote{Readers should note that the 2D projected half-light radius is theoretically smaller than the 3D half-light radius, $r_{1/2}$. \cite{2010MNRAS.406.1220W} show that $r_{1/2}/R_{e} \simeq 4/3$ for most surface brightness profiles with spherical symmetry. However, this difference is reduced in face-on views of flattened shapes \citep[i.e., $c/a < 1$ when $c$ is the length of the projection axis;][]{2022A&A...665A.159P}. The differences between $R_e$ and $r_{1/2}$ in this work are primarily induced by including observational effect.} Dust extinction reduces the central luminosity in the inner core, effectively increasing $R_e$. 

Scattered stellar emission smooths the central luminosity peak. This scattering effect can be roughly estimated by comparing the effective radius derived from direct (unscattered) sources to that derived from scattered sources, as provided by the \texttt{SKIRT} outputs. The effect of scattering is negligible compared to the impact of dust extinction. Finally, secondary emission from the gaseous disk may also influence the luminosity distribution, but this has little effect under our broad filter choices, which approximate the $V$ band in the rest frame (see Table \ref{tab:radii}). The effective radius generally decreases with longer filter wavelengths; for instance, a size reduction of approximately 25\% is observed with F400W at $z=2.8$, where dust emission remains negligible. Nevertheless, the effective radii remain significantly larger than $r_{1/2}$. 

The left panel in Figure \ref{fig:shape} shows the distribution of shape parameters at $z=2.8$. Half of the galaxies are classified as spheroidal shapes, while three galaxies are classified as oblate with $a \sim b > c$. 
We follow the criteria from \cite{2019MNRAS.484.5170Z} to classify their shapes, with boundaries represented by grey lines. The middle panel in Figure \ref{fig:shape} shows the size--mass relation of galaxies at $z=2.8$, characterized by $R_e$. The shapes of the galaxies largely agree with observations at $2.5 < z < 3$, although \textsc{Art-I}, \textsc{Enzo}, and \textsc{Changa} exhibit slightly higher $c/a$ values, indicating more spherical shapes compared to observed galaxies. Moreover, the size--mass relation of the galaxies shows strong agreement with observational data, and good inter-code convergence in size is achieved at $z=2.8$. Comparing the \textit{CosmoRun} galaxies with the median value of stellar mass and size of MW progenitors (yellow star in the middle panel in Figure \ref{fig:shape}), our results tend to overestimate both size and stellar mass, which is much mitigated at later epoch (see Figure 4 in Paper IV). 

The shape and size of galaxies are significantly altered by the inclusion of dust attenuation. Open markers in the panels represent the shape and size of the stellar distribution without the gas medium.\footnote{The shape of the stellar ellipsoid without the gas medium is measured by the eigenvalues of the inertia tensor. See \cite{2017MNRAS.472.1163C} for a detailed methodology.} The size, $b/a$, and $c/a$ increase with the inclusion of dust attenuation because the shape of the galaxy without dust attenuation represents the shape of the inner core, as small as $r_{1/2} \lesssim 0.5 \kpc$ in extreme cases of strong wet compaction. The inner core tends to have a prolate shape, as reflected in the open markers in the left panel, but could not fully represent the stellar distribution of the outer disk. On the other hand, $r_{1/2}$ in the codes with a weak compaction signature shows comparable values to the size of the gaseous disk. The stellar distribution in these three codes---\textsc{Enzo}, \textsc{Ramses}, and \textsc{Gear}---is largely prolate in the absence of dust attenuation and dark matter dominated (i.e., the dark matter mass within the $r_{1/2}$ sphere is larger than the stellar mass) at $z=2.8$, similar to the findings of \cite{2015MNRAS.453..408C}. See Appendix \ref{sec:transparent} for the mock snapshots and shape of the galaxies without dust attenuation.

We check whether the agreement in the size--mass relation with observations still holds at $z=1$ (right panel in Figure \ref{fig:shape}). The sizes of the \textit{AGORA} galaxies are consistent with the range of observational data presented in \citep{2024A&A...682A.110B}. We also note that the wide differences in size between cases with and without dust attenuation lessen at $z=1$, due to a combination of two effects: the decrease in disk cold gas at $z=1$ and the increase in $r_{1/2}$ from inside-out star formation. 

\begin{figure*}
    \centering
    \includegraphics[width=1.02\linewidth]{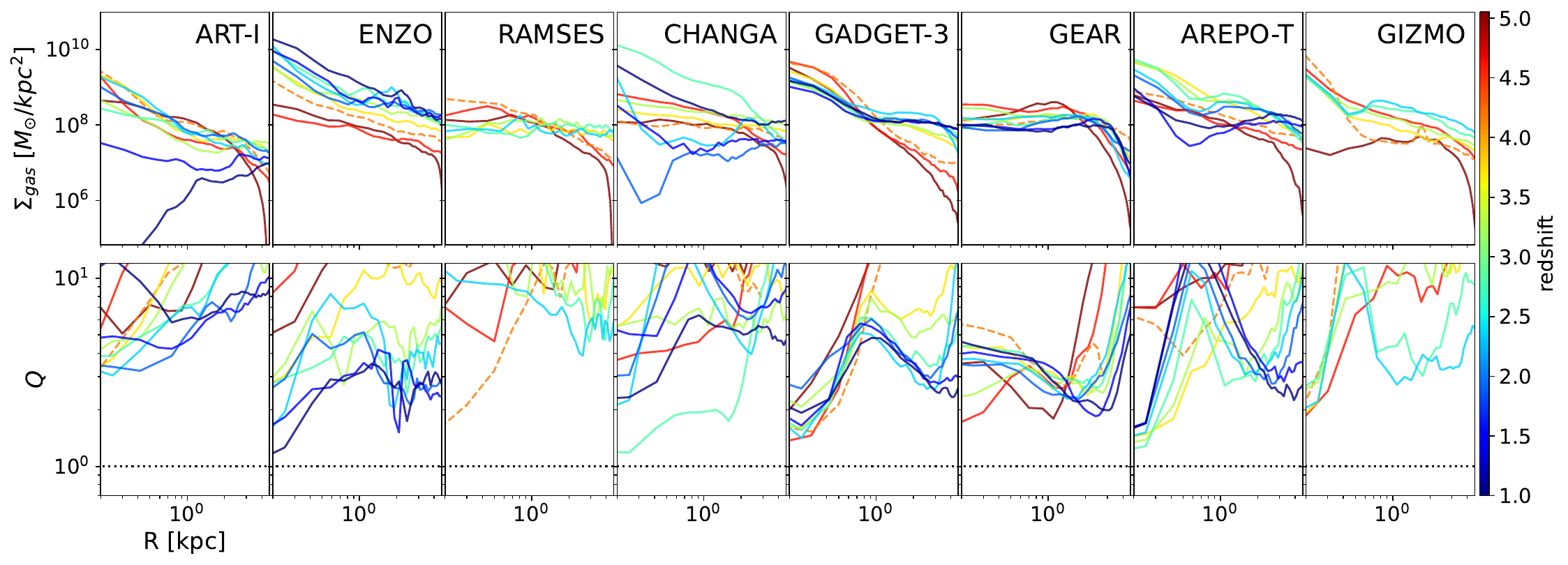}
    \caption{Radial profiles of the cold gas surface density ($T < 1.5 \times 10^4 \, \text{K}$; {\it upper panels}) and the \my{$Q$ stability} parameter ({\it lower panels}; averaged in annular bins). 
    The line colors represent redshifts, ranging from $z=5$ to $z=1$. 
    The {\it dashed lines} in each panel indicate the period that is heavily influenced by the major merger at $z \sim 4.5$. \textsc{Enzo} generally exhibits the lowest \my{$Q$} within $5 \kpc$, dropping to around 4 across all radii after $z=3$.
    For detailed discussions, see Section \ref{sec:dis_compaction}.}
    \label{fig:toomre}
    \vspace{3mm}
\end{figure*}

Finally, the Sérsic index of the galaxies, which range from $n=0.6$ to $1.1$ at $z=2.8$ and $n=0.5$ to $1.1$ at $z=1$, indicate that these systems are predominantly disk-dominated, consistent with findings from previous sections. The minimal growth of the Sérsic index over cosmic time at $z>1$ aligns with the assembly history of MW analogues \citep{2024arXiv241207829T}. The only exception is \textsc{Art-I}, which exhibits a higher Sérsic index of $n=2.5$ at $z=1$, indicating the possible presence of a strong bulge at that redshift. The period of quenched star formation at $0.8<z<1.6$ in \textsc{Art-I} results in a deficit of young stars in the disk plane, explaining this outcome (see Figure 7 in Paper IV for the star formation rate as a function of redshift, or Figure \ref{fig:toomre} for the radial profile of the surface density of cold gas). 


Overall, we compare the shape and size of galaxies at $z=2.8$ and $z=1$, showing good convergence among the participating codes and agreement with observed galaxies. However, this agreement, especially at $z=2.8$, is only achieved when dust attenuation is carefully considered. We present the same comparisons at $z=4.6$ and $z=3.5$ in Appendix \ref{sec:obser_highz}. In short, the \textit{CosmoRun} generally exhibits a spheroidal shape at those redshifts, while observations predict a prolate shape.

\vspace{3mm}

\section{Discussion}\label{sec:discuss}

\subsection{Why Haven't Some Galaxies Undergone Compaction?}\label{sec:dis_compaction}

We have classified the participating codes into two groups based on whether compaction is observed in the $M_{\star}-r_{1/2}$ space. \my{For a detailed discussion of the differences between the two groups, we present the radially averaged cold gas surface density and $Q$ stability parameter in Figure \ref{fig:toomre}. All participating codes show $Q > 1$ throughout cosmic time, suggesting that the disks remain globally stable against axisymmetric collapse, and that merger-driven inflows---rather than disk instability---play the dominant role in triggering compaction.}

We note that \textsc{Ramses} and \textsc{Gear}, the two codes that exhibit the least compaction (see Section \ref{sec:disk_growth}), utilize delayed cooling implementations in their supernova prescriptions. Previous studies, including the \textsc{Ramses} stellar feedback routine adopted in the {\it CosmoRun}, have found that delayed cooling inhibits the formation of dense star-forming clumps and disk structures in the galaxy's center \citep{2015MNRAS.452.1502D, 2018MNRAS.478..302S}. Paper III noted that employing a delayed cooling strategy can result in the accumulation of warm-hot gas in a dense state around star-forming regions. \textsc{Ramses} and \textsc{Gear} implement delayed cooling times of $10$ and $5 \Myr$, respectively, which are expected to prevent the concentration of baryonic mass in the inner regions, leading to thicker disk structures. The effect of these stellar feedback schemes is evident in the gas distribution within the galaxies, as shown in Figure \ref{fig:toomre}. In these codes, the radial profiles of cold gas surface density show minimal impact from the major merger at $z \sim 4.5$ (upper panels in Figure \ref{fig:toomre}). However, notable discrepancies exist between \textsc{Ramses} and \textsc{Gear} (see Table 3 in Paper IV and Figure 2 in Paper VI for galaxy gas masses in the \textit{CosmoRun}). The galaxy in \textsc{Gear} remains the most gas-rich at $z > 3$, whereas \textsc{Ramses} exhibits significantly lower gas mass. Consequently, \textsc{Ramses} displays a \my{$Q$} value that is more than twice that of \textsc{Gear} (see lower panels in Figure \ref{fig:toomre}). As shown in Figure \ref{fig:disk_decomp_z28}, \textsc{Ramses} displays an irregular gas distribution, while \textsc{Gear} shows a more spherical configuration. Insights from Paper VI indicate that \textsc{Gear} is the least efficient at driving outflows, whereas \textsc{Ramses} generates very strong gas outflows driven by stellar feedback. 

\textsc{Gadget-3} also implements the delayed cooling method but uses a cooling time based on the hot phase duration of SN feedback, $t_{\text{hot}} \propto n_0^{0.27} P_{0}^{-0.64}$, where $n_0$ is the ambient hydrogen density and $P_0$ is the ambient gas pressure \citep{2019MNRAS.484.2632S}. As a result, warm-hot gas in a dense state in the \textsc{Gadget-3} has an effectively low delayed cooling time with $t_{\text{delay}} \lesssim 10^5 \yr$. These gas particles can quickly cool and accrete into the dense disk. Indeed, the target galaxy in \textsc{Gadget-3} shows no clear indication of a contraction phase in terms of stellar size, exhibiting sizes approximately 0.5 dex smaller at $M_{\star} < 10^9\msun$, or at redshifts $z>5$ (see Figure \ref{fig:disk_evol} and \ref{fig:Mstar-size}). Consequently, the galaxy formed a rotationally supported disk at an earlier epoch, with $\Sigma_{{\star}, 1 \kpc} > 10^{9} \msun/\mathrm{kpc}^2$ as early as $z=5$. The \textsc{Gadget-3} code group speculates that relatively low supernova feedback energy may have contributed to the early accumulation of gas content by $z>5$.

Wet compaction is also less pronounced in \textsc{Enzo}, which does not implement delayed cooling. Notably, \textsc{Enzo} has the highest gas mass on galaxy scales at $z=2.8$, along with \textsc{Gear}, with values more than twice those of the other two AMR codes (see Table 3 in Paper IV and Figure 2 in Paper VI). Paper VI speculates that the strong, purely thermal feedback in \textsc{Enzo} may be linked to the higher gas mass and SFR at $z\lesssim 3$. As a result, \textsc{Enzo} generally shows the lowest $Q$ within $5 \kpc$, reaching $Q \sim 4$ at all radii after $z=3$. We examine the two-dimensional distribution of \my{$Q$ stability parameter} in \textsc{Enzo}, which shows $Q < 1$ inside and around the clumps and $Q \sim 2$–$3$ in the proto-clump region (not shown). The formation of clumps can still be driven, especially at high redshift, by various scenarios such as compressive turbulence \citep{2016MNRAS.456.2052I, 2025MNRAS.538L...9M, 2025arXiv250107097G}, although $Q < 1$ is traditionally considered a standard condition for instability. This suggests that as the gas-rich disk forms multiple massive clumps, wet compaction and subsequent inside-out quenching may be obscured.

\begin{figure*}
    \centering
    \includegraphics[width = 1.02\linewidth]{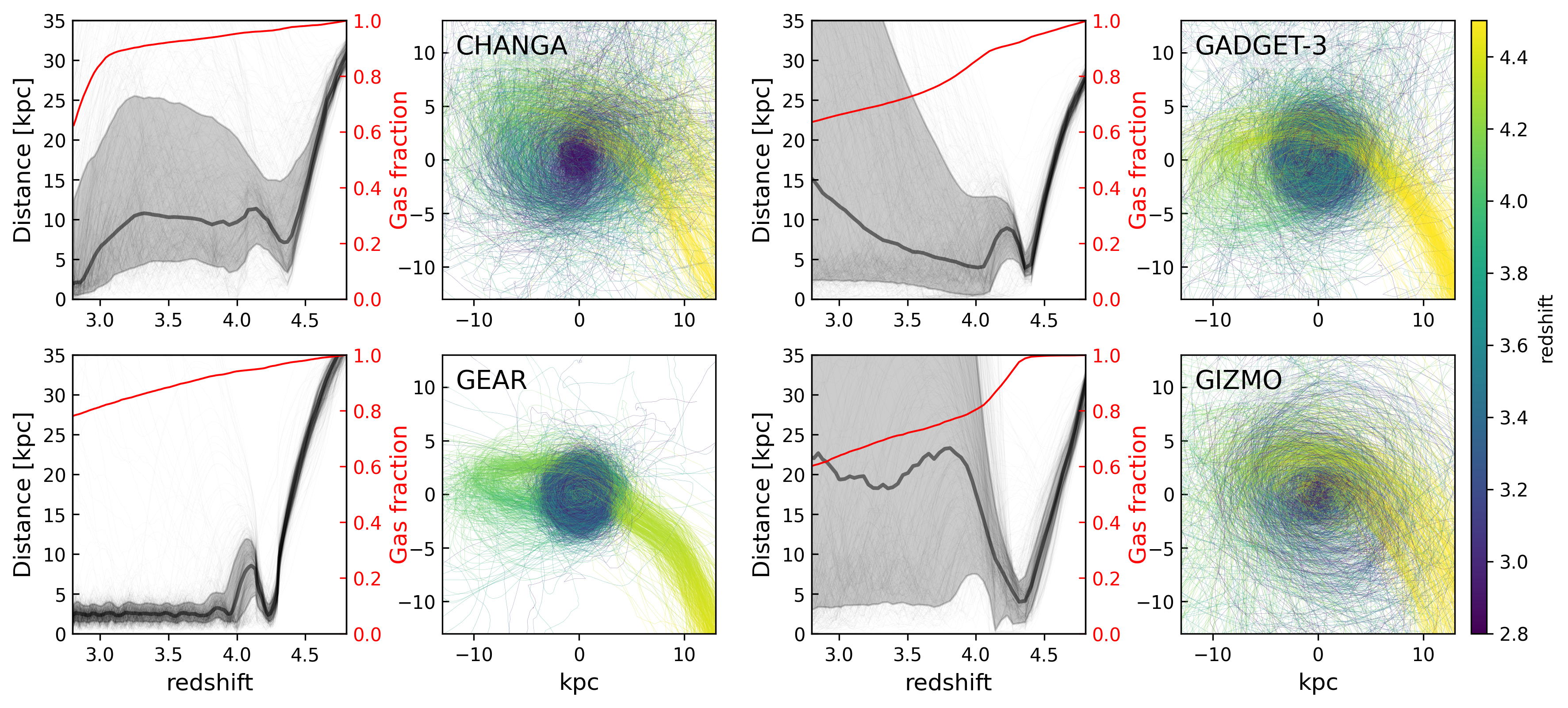}
    \caption{The trajectories of gas particles of the incoming merging satellites during and after the (major) merger event at $z\sim4.5$, for the four particle-based codes (\textsc{Changa}, \textsc{Gadget-3}, \textsc{Gear}, and \textsc{Gizmo}). 
    Each line in the {\it left panels} represents the particle's 3D distance from the target galaxy's center, and the {\it thick black line} indicates the median.   
    The {\it gray shaded region} show the 16\%--84\% range. 
    The fraction of particles that remain in the gas phase at each redshift (i.e., not converted into star particles) is shown as a {\it thick red line}. 
    The {\it right panels} show the trajectories at a face-on angle for 300 randomly selected particles, with its line color slowly changing to indicate the redshift. 
    The influx of gas particles in \textsc{Changa} is initially limited to the circumgalactic region, mostly with distances $>5 \kpc$. 
    Only at $z\sim3$ does the gas flow into the center of the target galaxy, inducing a starburst.
    In contrast, the gas particles in \textsc{Gear} reach deeper into the center of the target galaxy, already at $z\sim4$.  
    For detailed discussions, see Section \ref{sec:dis_timediff}.}
    \label{fig:gas_track}
    \vspace{2mm}
\end{figure*}

\subsection{Differences In Compaction Timescales}\label{sec:dis_timediff}

We have shown that the primary source of compaction in most codes is the wet merger at $z \sim 4.5$, which serves as a gas supply and induces a loss of angular momentum, resulting in inward accretion. \textsc{Arepo-t} and \textsc{Gizmo} exhibit the most pronounced reactions to the merger, resulting in the rapid formation of an inner core. However, a time difference of about $300 \Myr$ in the compaction event is observed between these two codes. Notably, \textsc{Gizmo} exhibits a lower $Q$ and a higher gas surface density in the inner region than \textsc{Arepo-t} at $z=4.6$, just before the major merger. This is likely induced by gas fragmentation at $z=5$ in \textsc{Gizmo}, making it easier for it to undergo \textit{violent} disk instability that drives gas into the center \citep{2014MNRAS.438.1870D, 2023MNRAS.522.4515L}. This discrepancy could stem from variations in the timing of starbursts during the early and late stages of mergers, a topic that will be explored further in Paper IX.

\textsc{Changa} presents a particularly interesting case, where similar inner core growth is observed but with an $800 \Myr$ delay between the major merger and wet compaction. Rejuvenation events occur at $z=2.8$ and $z=1.2$ (see Figure \ref{fig:mean_radius} and Section \ref{sec:sfh}). A similar but weaker trend is observed at $z=4.4$, where the star formation rate rises due to a major merger but subsequently declines and plateaus until $z=3$, accompanied by a slight decrease in the mean radii of stellar particles around $z \sim 4.5$. 

We trace the gas particles of the merging satellites in Figure \ref{fig:gas_track} and show the trajectories of gas during and after the major merger at $z \sim 4.5$.\footnote{The results from the AMR codes and \textsc{Arepo-t} are omitted, as tracing gas over multiple timesteps in these codes is difficult to achieve without the implementation of tracer particles \citep[e.g.,][]{2013MNRAS.435.1426G, 2019A&A...621A..96C}.} In \textsc{Changa}, gas particles exhibit greater distances at the first pericenter ($z=4.3$) compared to the other codes (left panels of each code). After $z=4$, more than 90\% of the inflowing gas particles in \textsc{Changa} remain as gas, whereas \textsc{Gadget-3} and \textsc{Gizmo} show that more than 20\% of the gas is converted into stellar particles after the major merger, accompany with the wet compaction. Only after $z=3$ does the gas flow into the galactic region, leading to a rapid starburst. During the major merger, inflowing gas fails to efficiently accrete into the central region. Instead, it is rapidly expelled into the circumgalactic region by supernova feedback. The ``superbubble" model in \textsc{Changa}, presented by \citep{2014MNRAS.442.3013K}, which drives strong outflows, explains this phenomenon. 

These findings highlight the intricate interplay among feedback mechanisms, disk instabilities, and merger dynamics in determining compaction timescales and gas accretion processes across codes. A detailed exploration of inflow/outflow dynamics and cold gas content within the disk may provide further insights into these observed delays. A more comprehensive study on the connection between mergers and starbursts will be presented in Paper IX.

\subsection{Limitations \& Caveats}
\my{We primarily attribute the inter-code differences in disk formation to the choice of stellar feedback prescriptions. However, we also find a possible dependence on the hydrodynamic solver in the strength of wet compaction.} \del{Finally, we note that }\my{The} AMR codes generally exhibit weaker compaction than SPH and hybrid codes, although morphological features among the three AMR codes --- \textsc{Art-I}, \textsc{Enzo}, and \textsc{Ramses} --- vary significantly. This may be explained by findings from \cite{2023ApJ...950..132H}, which suggest that gas in Lagrangian codes collapses to much higher densities than in Eulerian codes, where stability is maintained by the minimal cell size. \my{Yet, the differences between the hydrodynamic solvers are not significant enough to make a strong statement.}

Additionally, it is important to emphasize that such discrepancies arising from different hydrodynamic solvers might become more significant at high redshift due to limited numerical resolution. For instance, wet compaction in SPH codes could be artificially exaggerated at insufficient resolution owing to excessive angular momentum transfer \citep[e.g.,][]{2000ApJ...538..477N}. In principle, such limitations could hinder studies of rotationally supported disks forming as early as $z\sim8$, as reported in numerical simulations \citep{2022ApJ...928..106T}. However, the onset of disk formation in Milky Way–mass systems remains uncertain and depends sensitively on the mass accretion history \citep{2024A&A...683A.236P}. In our \textit{CosmoRun} halo, all participating codes show a rapid kinematic transition only at $z<5$, with little evidence for an earlier compact spin-up of gas that higher resolution might otherwise reveal. Thus, for this specific accretion history, the earliest disk appears after $z\sim5$, and the disk formation epoch we report at $z\sim4$ should be regarded as reliable given our adopted resolution.


\section{Summary and Conclusion}\label{sec:conclusion}  

In this work, we use cosmological zoom-in simulations of a Milky Way-mass halo from eight codes in the \textit{AGORA} project to investigate the formation and properties of galactic disks across a redshift range of $z \sim 1-5$. Target redshifts of $z=2.8$ and $z=1$ are chosen to minimize the impact of discrepancies in merger timing. Thin disk, thick disk, and spheroid stellar components are identified based on the kinematic properties of each star. The key findings of this work are summarized below:
\begin{itemize}
	\item We observe the rapid formation of compact cores with rotation-dominated kinematics for some codes, notably on \textsc{Art-I}, \textsc{Changa}, \textsc{Gadget-3}, \textsc{Arepo-t}, and \textsc{Gizmo} (Figure \ref{fig:disk_evol}). The formation of the core coincides with an increase in the disk-to-total ratio (D/T) and the stellar surface density within a 1 kpc radius ($\Sigma_{\star, 1,\text{kpc}}$), as well as a decrease in the half-mass radius ($r_{1/2}$), although these connections are less pronounced in \textsc{Gadget-3}.
	\item Wet compaction is highlighted in the evolution of size ($r_{1/2}$) with stellar mass (Figure \ref{fig:Mstar-size}), where \textsc{Art-I}, \textsc{Changa}, \textsc{Gadget-3}, \textsc{Arepo-t}, and \textsc{Gizmo} show contraction and expansion phases in size. The other codes --- \textsc{Enzo}, \textsc{Ramses}, \textsc{Gear} --- exhibit weaker signatures of compaction. \my{The ratio of radial velocity dispersion to tangential velocity, $\sigma_r / V_{\phi}$, in the stellar component supports this classification: in codes with a strong signature of compaction, the stellar component exhibits a rapid transition to $\sigma_r / V_{\phi} \gtrsim 2$, whereas a similar rise in the gas component is observed across all codes (Figure \ref{fig:vcirc-sigma}). }
	\item The thin disk contains a younger stellar population than the thick disk and spheroid components (Figure \ref{fig:age_histogram}). Additionally, the age gradients of the thin disk components are smaller or negative compared to the spheroid component across all \textit{AGORA} codes. The codes with clear wet compaction (\textsc{Art-I}, \textsc{Changa}, \textsc{Gadget-3}, \textsc{Arepo-t}, and \textsc{Gizmo}) display negative total age gradients. An inside-out pattern of star formation is observed, where the mean radius of newly formed stars increases over time (Figure \ref{fig:mean_radius}), particularly in the thin disk component. Stars in these codes generally show minimal radial migration. 
	\item Codes with weak compaction signatures (\textsc{Enzo}, \textsc{Ramses}, \textsc{Gear}) exhibit positive or flat age gradients in the stellar disk. The primary source of positive gradients is outward migration, which reaches scales of up to $\sim 2 \kpc$ in \textsc{Art-I}, \textsc{Ramses}, and \textsc{Gear}, and even higher in \textsc{Enzo}. The absence of a steep central gravitational potential in \textsc{Art-I}, \textsc{Ramses} and \textsc{Gear} facilitates outward migration (Figure \ref{fig:v_circ}). In \textsc{Art-I}, strong inside-out quenching counteracts the impact of outward migration on the overall stellar age gradient.
	\item The comparison of galaxy shapes and sizes at $z=2.8$ and $z=1$ (Figure \ref{fig:shape}) shows good convergence among participating codes and alignment with observations. However, this agreement, especially at $z=2.8$, is only achieved when dust attenuation is carefully considered. 
	\item Supernova feedback with delayed cooling in \textsc{Ramses} and \textsc{Gear} (10 and 5 Myr, respectively) prevents the galaxies in these codes from undergoing rapid compaction. Those codes exhibit lower inner gas surface density. The formation of multiple clumps in \textsc{Enzo}, indicated by $Q \sim 4$ at all radii after $z=3$, obscures wet compaction and contraction in size (Figure \ref{fig:toomre}).
	\item Differences in the timing of wet compaction onset are observed between the codes, depending on the phase of the major merger that triggers the starburst. In \textsc{Changa}, an $800 \Myr$ delay between the major merger and compaction is observed, caused by feedback-driven outflows that expel gas into the circumgalactic region, where it later cools and accretes (Figure \ref{fig:gas_track}). 
\end{itemize}

\del{We primarily attribute the inter-code differences in disk formation to the choice of stellar feedback prescriptions. However, we also find a possible dependence on the hydrodynamic solver in the strength of wet compaction.} The \textit{AGORA} collaboration is working to incorporate new codes as well as simulations with alternative feedback schemes, such as \textsc{Swift} and \textsc{Ramses-vintergatan}. Higher-resolution runs within \textit{AGORA} are also already underway. These additions will help test convergence across different hydrodynamic schemes and further investigate potential discrepancies in disk formation based on hydrodynamic solvers.

Our future goals include studying the formation of thin disks and the secular evolution of spiral arms and bars at $z<1$, stemming from the distinct characteristics of the thick disk and inner core in the \textit{CosmoRun} simulations.

\begin{acknowledgments}
We thank Jeong Hwan Lee for his guidance in performing the \texttt{galfit} analysis. J.-H.K.'s work was supported by the National Research Foundation of Korea (NRF) grant funded by the Korea government (MSIT) (No. 2022M3K3A1093827 and No. 2023R1A2C1003244). 
His work was also supported by the National Institute of Supercomputing and Network/Korea Institute
of Science and Technology Information with supercomputing resources including technical support, grants KSC-2021-CRE-0442, KSC- 2022-CRE-0355 and KSC-2024-CRE-0232. His work was also supported by the Global-LAMP Program of the NRF grant funded by the Ministry of Education (No. RS-2023-00301976). R.R.C. acknowledges financial support from GMV; from the Spanish Ministry of Science and Innovation through the research grant PID2021-123417OB-I00, funded by MCIN/AEI/10.13039/501100011033/FEDER, EU; and from the IND2022/TIC-23643 project funded by the Comunidad de Madrid. HV was supported by DGAPA-PAPIIT-UNAM project IN111425. ChaNGa runs were done in the CNS-IPICYT-SECIHTI,  LNS-SECIHTI del Sureste and LAMOD-UNAM. DC is supported by the Ministerio de Ciencia, Innovación y Universidades under research grants PID2021-122603NB-C21 and CNS2024-154550.
KN is supported in part by the MEXT/JSPS KAKENHI grant number 22K21349, 24H00002, and 24H00241. KN also acknowledges the support from the Kavli IPMU, World Premier Research Center Initiative (WPI), UTIAS, the University of Tokyo. 
\end{acknowledgments}
\software{yt \citep{2011ApJS..192....9T}, pytreegrav \citep{Grudić2021}, agora\_analysis \citep{2024ApJ...962...29S}} 
\bibliography{main}{}
\bibliographystyle{aasjournal}

\appendix

\section{Kinematic decomposition at $z=1$}\label{sec:z1}

\begin{figure*}
    \centering
    \includegraphics[width = 0.96\linewidth]{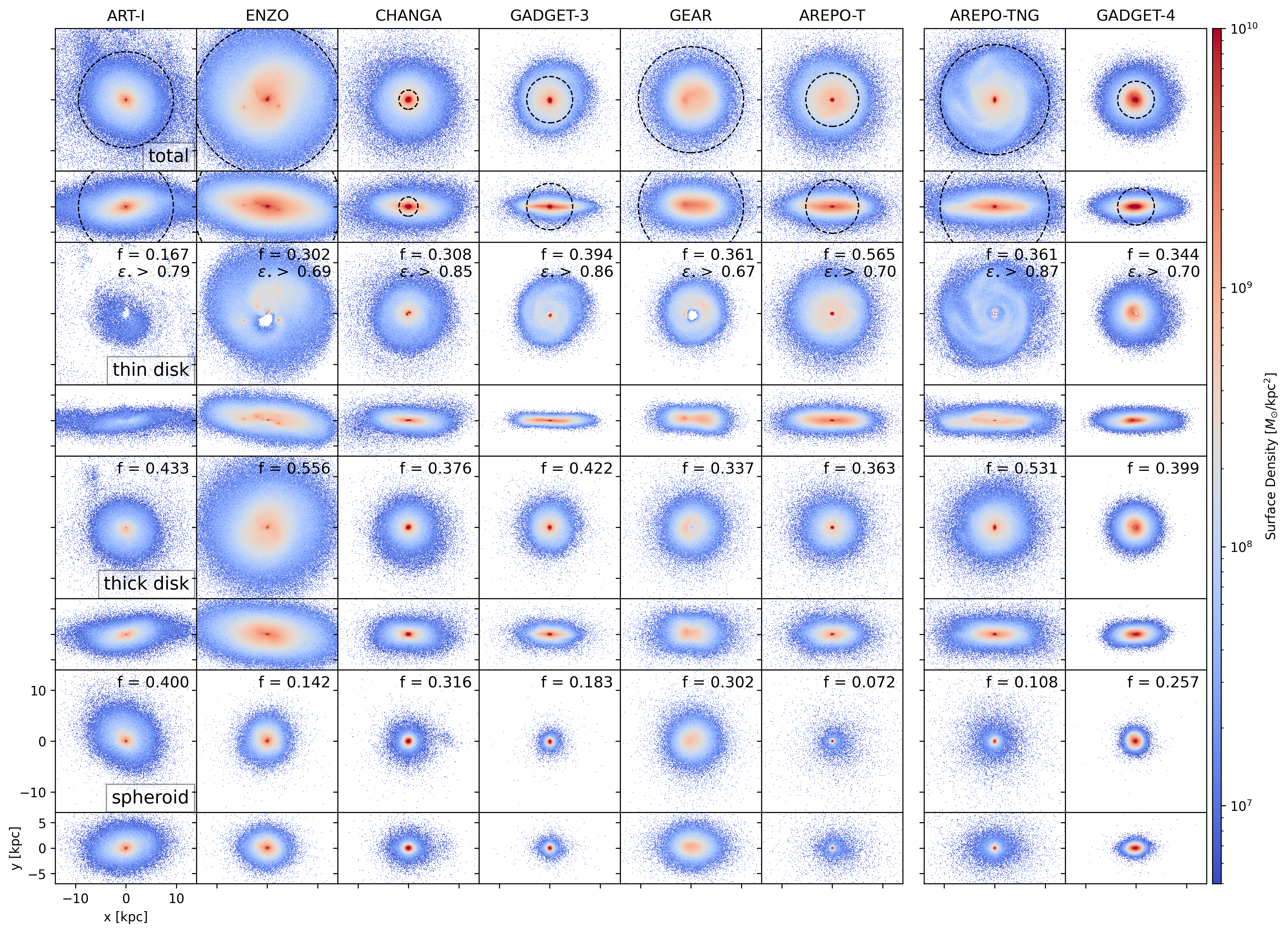}
    \caption{Same as Figure \ref{fig:mock_z28}, but at $z=1$.  
    Note that six out of eight codes seen in Figure \ref{fig:mock_z28} reach redshift $z = 1$, while \textsc{Ramses} and \textsc{Gizmo} end their simulations at $z=2$ (see Sections \ref{sec:suite} and \ref{sec:mock}). \my{Results from \textsc{Arepo-tng} and \textsc{Gadget-4} are also shown.}
    Note also that we present the $z=1.03$ snapshot for \textsc{Gear} to marginalize the effect of a minor merger event at $z=1$. 
    For more information, see Appendix \ref{sec:z1}.}
    \label{fig:disk_decomp_z1}
    \vspace{2mm}
\end{figure*}

Figure \ref{fig:disk_decomp_z1} presents face-on and edge-on views of the stellar surface density distributions at $z=1$, along with the results of stellar kinematic decomposition. The circularity threshold between the thin and thick disk, $\epsilon_{\text{thre}}$, in each code is larger than at $z=2.8$, except in \textsc{Enzo}. The thin disk fractions increase from $z=2.8$ to $z=1$, except in \textsc{Art-I}, which visually exhibits a disk warp at $z=1$. This indicates that thin disks in the \textit{CosmoRun} galaxies generally form at $z<2.8$, while the disk-to-total fraction (D/T) remains largely steady (see Section \ref{sec:disk_growth}). \my{The results from \textsc{Arepo-tng} and \textsc{Gadget-4} codes are also presented. See Appendix \ref{sec:tng} for more information of the codes. }

\begin{figure}
    \centering
    \includegraphics[width = \linewidth]{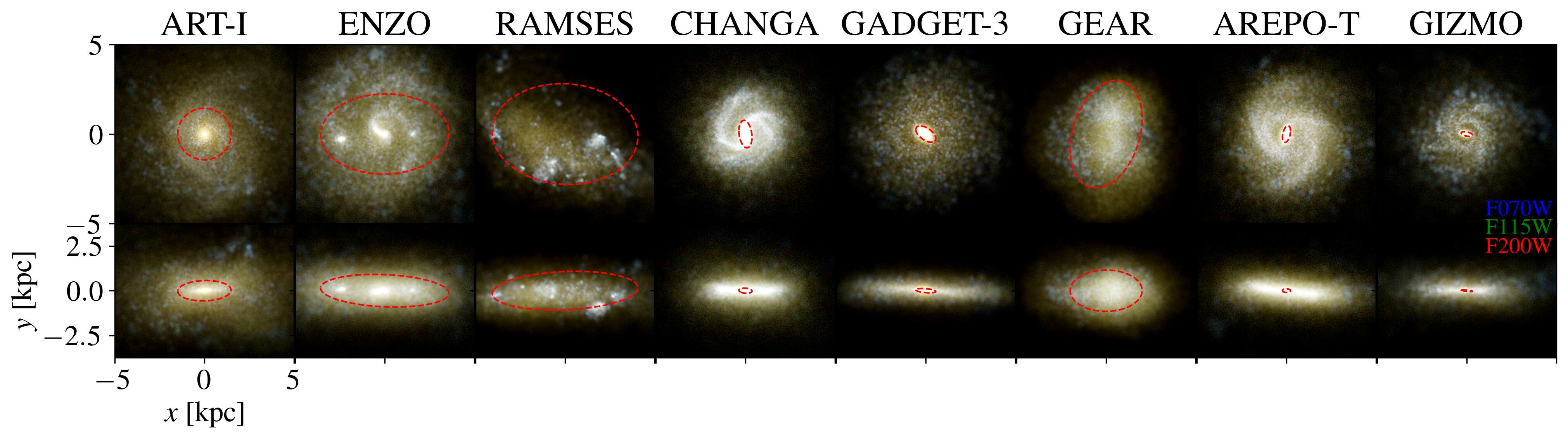}
    \caption{Same as Figure \ref{fig:mock_z28}, but without dust and gas medium. The red ellipse represents the stellar ellipsoid encompassing half of the total stellar mass.}
    \label{fig:mock_z28_trans}
    \vspace{2mm}
\end{figure}
\begin{figure}
    \centering
    \includegraphics[width = 0.86\linewidth]{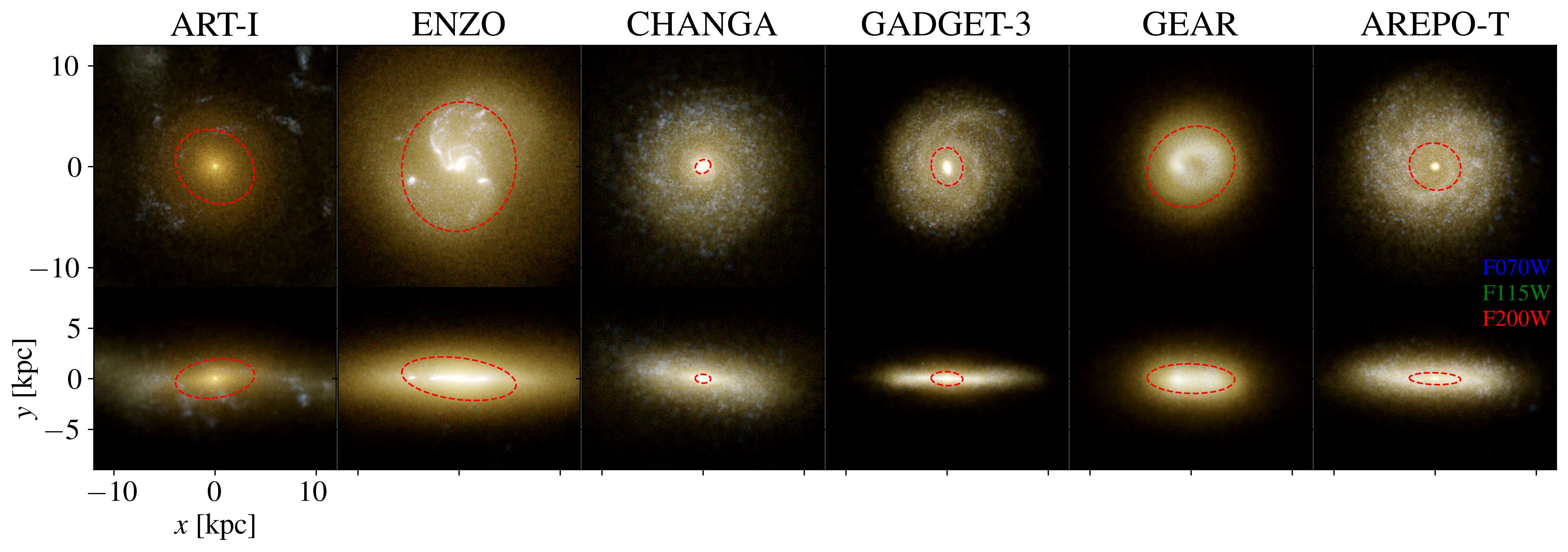}
    \caption{Same as Figure \ref{fig:mock_z1}, but without dust and gas medium. The red ellipse represents the stellar ellipsoid encompassing half of the total stellar mass.}
    \label{fig:mock_z1_trans}
    \vspace{2mm}
\end{figure}

\section{SKIRT mock images without dust}\label{sec:transparent}

We display the edge-on and face-on views of the mock snapshots in Figures \ref{fig:mock_z28_trans} and \ref{fig:mock_z1_trans}, but without dust and gas. The red ellipse represents the projection of the ellipsoid, where the directions of the semi-axes are the eigenvectors of the inertia tensor of the stellar mass distribution. The shapes of galaxies with a strong signature of compaction (\textsc{Art-I}, \textsc{Changa}, \textsc{Gadget-3}, \textsc{Arepo-t}, and \textsc{Gizmo}) follow the shape of their inner cores at $z=2.8$, which are generally prolate.

\section{Comparison With Observations at higher redshift }\label{sec:obser_highz}
\begin{figure}
    \centering
    \includegraphics[width = \linewidth]{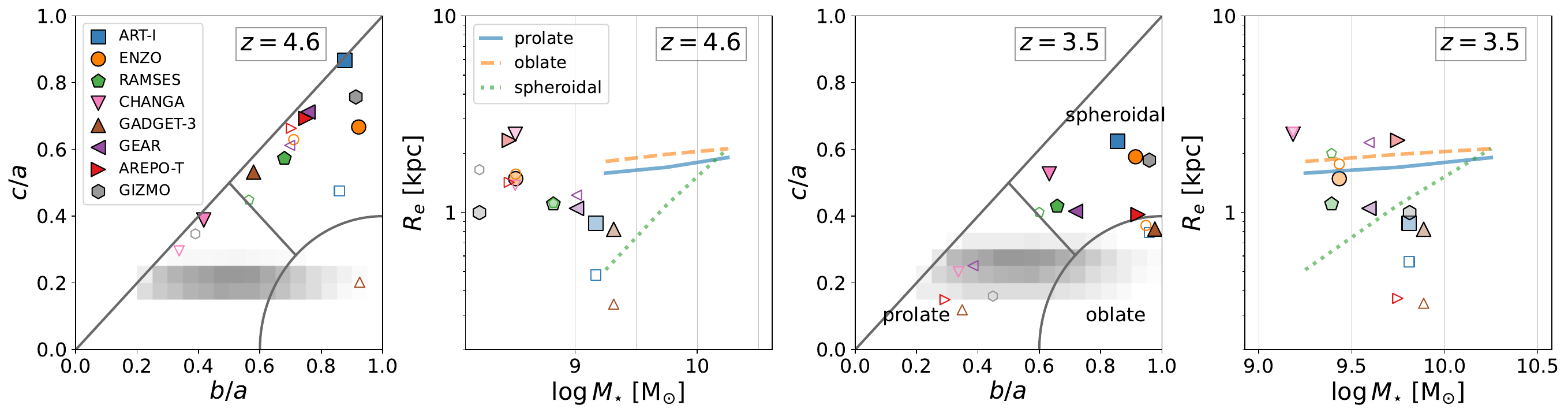}
    \caption{Similar to Figure \ref{fig:shape}, but at $z=4.6$ (\textit{left two panels}) and $z=3.5$ (\textit{right two panels}). The \textit{gray-shaded region} indicates the observational results at $z=3$--$8$ for $\log{(M_{\star}/\mathrm{M_{\odot}})} = 9.0$--$9.5$ (\textit{left}) and $\log{(M_{\star}/\mathrm{M_{\odot}})} = 9.5$--$10.0$ (\textit{right}) \citep{2024ApJ...963...54P}.
    }
    \label{fig:observation_z46_z35}
    \vspace{2mm}
\end{figure}

Figure \ref{fig:observation_z46_z35} shows the 3D axis ratios ($b/a$, $c/a$) and the size-mass relation at $z=4.6$ (just before the major merger at $z\sim4.5$) and $z=3.5$. The shape of galaxies at $z=4.6$ exhibits a spheroidal shape for most of the codes. While most galaxies at $z=4.6$ have lower stellar masses than the lowest stellar mass bin in \citet{2024ApJ...963...54P}, an inconsistency in the shape of galaxies between \textit{CosmoRun} and the observations is evident. The sizes of the galaxies show results comparable to observations, though there is some scatter between the codes.

Similar results are found at $z=3.5$: the galaxies tend to have a spheroidal shape, while observations predict a predominantly prolate distribution. The only exceptions are \textsc{Changa} and \textsc{Gear}, which match the observations in both size and shape when considering the shape without dust extinction effects. However, the shape shifts toward a more spheroidal morphology after accounting for dust extinction.

\section{Results from Gadget-4 and AREPO-TNG}\label{sec:tng} 

\begin{figure}
    \centering
    \includegraphics[width = 0.8\linewidth]{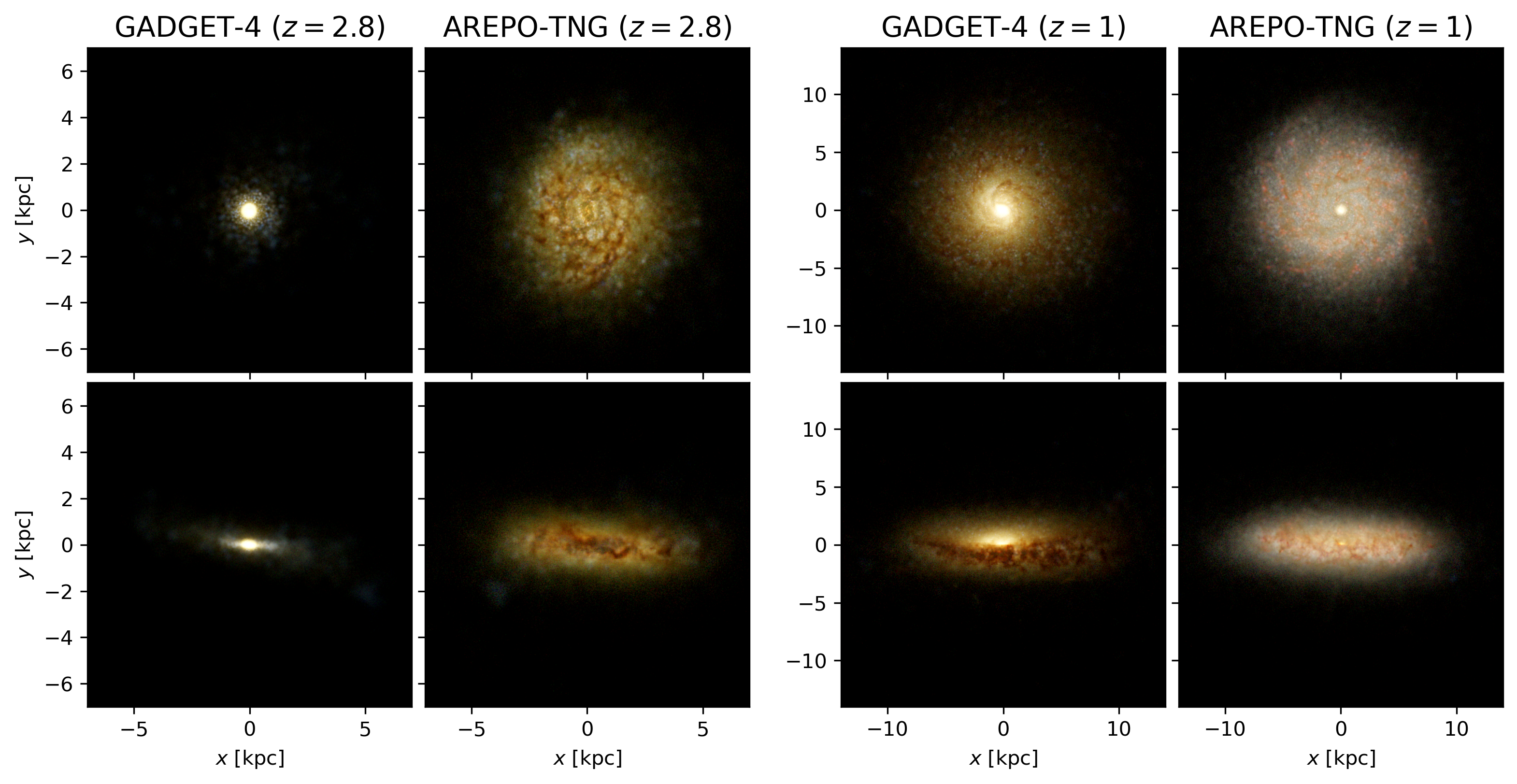}
    \caption{Same as Figures \ref{fig:mock_z28} and \ref{fig:mock_z1} but with \textsc{Gadget-4} and \textsc{Arepo-tng}. }
    \label{fig:snapshot_appendix}
\end{figure}

\begin{figure*}
    \centering
    \includegraphics[width = \linewidth]{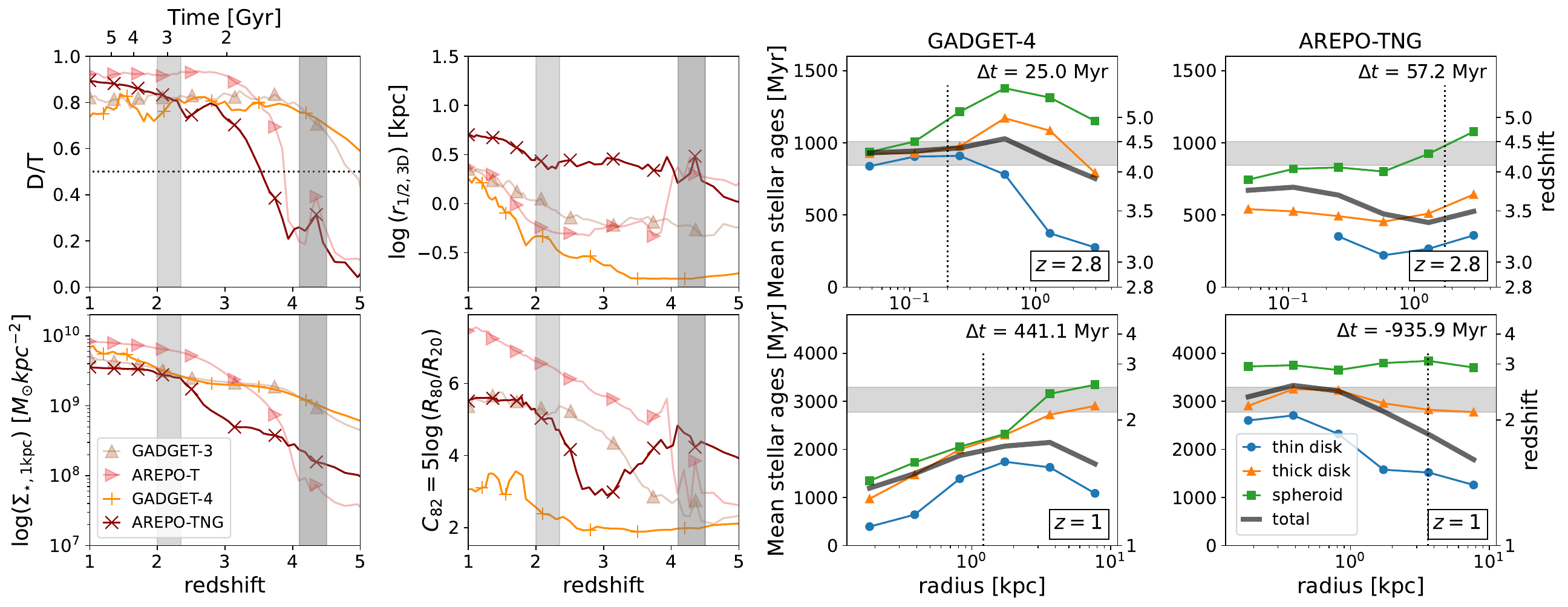}
    \caption{Same as Figures \ref{fig:disk_evol} and \ref{fig:age_histogram} but with \textsc{Gadget-4} and \textsc{Arepo-tng}. 
    For more information, see Appendix \ref{sec:tng}.}
    \label{fig:arepo-tng}
    \vspace{2mm}
\end{figure*}

\begin{figure*}
    \centering
    \includegraphics[width = \linewidth]{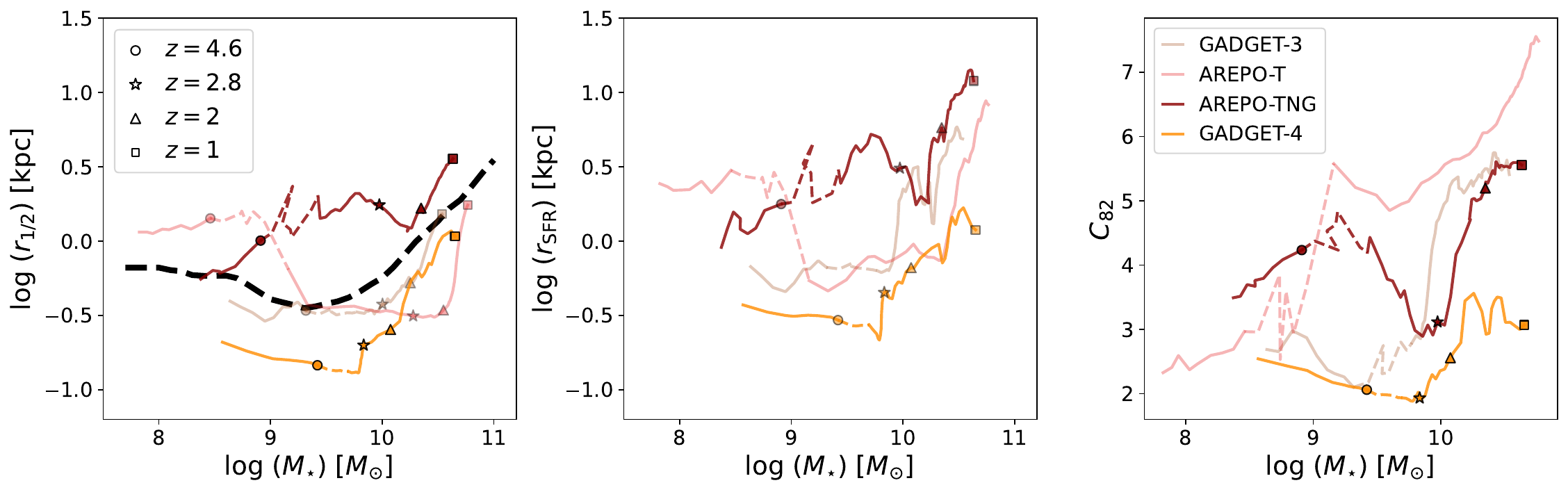}
    \caption{\my{Same as Figure \ref{fig:Mstar-size} but with \textsc{Gadget-4} and \textsc{Arepo-tng}. 
    For more information, see Appendix \ref{sec:tng}.}}
    \label{fig:mstar-size-app}
    \vspace{2mm}
\end{figure*}

All participating codes have been tested using the calibration steps described in Paper III to verify the convergence of shared physics (e.g., gas cooling and heating) and to assess the impact of code-dependent physics, particularly stellar feedback. \del{The \textsc{Gadget4-Osaka} code presented here differs from the \textit{CosmoRun} by including AGN feedback, employing different star formation thresholds, and using a different IMF.}  \my{The \textsc{Gadget-4} code presented here has newly been added to the AGORA \textit{CosmoRun} suite, with full calibration steps.} Meanwhile, \textsc{Arepo-tng} implements the physics models used in the \my{\textsc{IllustrisTng} simulation}, which have their own set of physics recipes (e.g., heating and cooling, star formation) distinct from those in the \textit{CosmoRun}, but with the AGN models turned off. \my{Figure \ref{fig:snapshot_appendix} displays mock snapshots of \textsc{Gadget-4} and \textsc{Arepo-tng}.} \del{Readers should note that the Chabrier initial mass function is also used for the mock visualization of these results, while \textsc{Gadget-4} implements a metallicity- and redshift-dependent top-heavy initial mass function.}

\my{Readers should note that \textsc{Gadget-4} tracks dust species within each gas parcel, which are used to generate mock snapshots (see Appendix \ref{sec:gadget4} for more details).} To generate the \texttt{SKIRT} mock images in \textsc{Arepo-tng}, we adopt a slightly different method, following \cite{2024A&A...683A.181B}, to estimate the dust content rather than using equation \ref{eqn_dust}. Specifically, gas is assumed to contain no dust if its temperature satisfies:

\begin{equation}
	\log{\left(\frac{T_{\rm gas}}{\mathrm{K}}\right)} > 6 + 0.25\log{\left(\frac{\rho_{\rm gas}}{10^{10} \mathrm{M_{\odot}} h^2 \mathrm{kpc}^{-3}}\right)}
\end{equation}

\my{Figures \ref{fig:arepo-tng} and \ref{fig:mstar-size-app} present} key results obtained with \textsc{Gadget-4} and \textsc{Arepo-tng}. \my{\textsc{Gadget-4} exhibits compact stellar sizes with higher D/T ratios, along with a flat stellar age gradient at $z=2.8$. Already at $z=5$, the galaxy reaches D/T $> 0.5$ with $r_{1/2} \sim 0.2 \kpc$ (Figure \ref{fig:arepo-tng}), which is more compact and has more rotationally dominated kinematics than the other codes. No apparent size contraction is observed during or after the major merger at $z=4.5$. Albeit more compact, these trends are similar to those of \textsc{Gadget-3}. The stellar age gradient is negative in the outskirts with $r > r_{1/2}$ at $z=2.8$ and $z=1$, while the positive gradient in the inner region suggests that the dense core is not completely quenched until $z=1$, and multiple compaction events could be triggered. The on-the-fly dust model in \textsc{Gadget-4} allows us to compare its dust content with that obtained using a fixed dust-to-metal ratio of 0.2. At $z=2.8$, the total dust mass is about two dex lower than in the fixed ratio model, which is likely due to the metallicity- and density-dependent dust growth model \citep{2017MNRAS.466..105A,2020MNRAS.491.3844A} in \textsc{Gadget-4}. This discrepancy disappears by $z=1$, where the two methods converge within 0.11 dex. However, differences in the spatial distribution remain evident at $z=1$: the on-the-fly model shows higher central brightness (corresponding to less dust) compared to the fixed dust-to-metal model (not shown). The elevated star formation rate at $z=1$ suggests that supernova-driven dust destruction may be responsible for this difference \citep{2025arXiv250818374B}. }

\my{In contrast, \textsc{Arepo-tng} transitions into a rotationally dominated system at $z < 3.5$. The galaxy size in \textsc{Arepo-tng}, both $r_{1/2}$ and $r_{\rm SFR}$, increases monotonically with mild fluctuations (Figure \ref{fig:mstar-size-app}), in contrast to the evolution seen in \textsc{Arepo-t}. In \textsc{Arepo-tng}, a flat stellar age gradient is observed at $z=2.8$, while a negative age gradient, especially for the thin disk component, is present at $z=1$.}

The differences in disk formation patterns among these codes and the \textit{CosmoRun} underscore the various pathways to disk formation that arise from differing physical prescriptions. 

\section{CALIBRATION OF THE Gadget-4 CODE}\label{sec:gadget4}

Following the calibration procedure described in Paper III, and later adopted by \textsc{Arepo-t} in Paper IV, we present here the corresponding results obtained by the \textsc{Gadget4-Osaka} code group. We show the same figures as in our Paper III, now for a total of nine codes. Our Figures 20–28 and 31–32 correspond to Figures 2–4, 6, 11, 9, 12, 14, 20, 19 and 22 of Paper III, in that order. Figure \ref{Ap:Cal4_2} corresponds to the $z = 4$ panels across Figures 15–16, 18, and 21 in Paper III. We do not repeat here the full details and discussion regarding the observed differences between the codes; for that, we refer the reader to Sections 5 and 6 of Paper III. In line with what was done for all other participating codes in Section 3.2 of Paper III, here we briefly describe the physics included in the \textsc{Gadget4-Osaka} final runs. For a complete overview of how each code in \textit{CosmoRun} operates, we direct the reader to our earlier papers — Paper I for gravitational dynamics, Paper II for hydrodynamics, and Paper III for the cosmological simulation specifications and calibration methodology.

\subsection{Code-dependent Physics In \textsc{Gadget-4}} \label{sec:g4phys}

\textsc{Gadget4-Osaka} is a modified version of the N-body/SPH code \textsc{Gadget-4} \citep{Springel_2021}, enhanced with a suite of subgrid physics modules for star formation and feedback developed by the Osaka group. Gravitational forces are computed using a Fast Multipole Method-Particle Mesh (FMM-PM) scheme. The hydrodynamics are computed using the pressure-energy formulation of SPH \citep{Hopkins_2012,Saitoh_2013}, implemented with the Wendland C4 kernel \citep{Dehnen_2012}, $128 \pm 4$ neighbors, and a time-step limiter \citep{Saitoh_2009}. Radiative cooling and primordial chemistry are computed by the \textsc{Grackle-v3.3.1} library.

The stellar feedback prescription follows the framework of \citet{Oku_2022,Oku2024}, which distinguishes between a local mechanical feedback model and a hot galactic wind component. The mechanical feedback injects the terminal momentum from unresolved supernovae remnants, calibrated using high-resolution simulations of superbubbles.  The galactic wind properties are informed by the TIGRESS framework \citep{Kim_2020}, with launched wind particles temporarily decoupled from hydrodynamic forces to mimic free-streaming escape from dense star-forming regions. 

For the AGORA Cal-4 run, the total energy from Type II and Type Ia supernovae is boosted to $E_{SN} = 10 \times 10^{51}$ ergs/SN to satisfy the stellar mass requirement at $z=4$. To ensure compatibility with the AGORA framework, this run adopts a fixed Chabrier IMF, with metal yields computed using the \textsc{CELib} library \cite{Saitoh_2017}, rather than the metallicity- and redshift-dependent top-heavy IMF employed in the fiducial CROCODILE simulations \citep{Oku2024}. Specifically, for SNII feedback, we combine yield tables from \cite{Portinari} and \cite{Nomoto}. This choice yields a relatively high effective metal yield ($0.042$) compared with the $0.025$ in \textsc{Gadget3-Osaka}; the discrepancy disappears if we instead use only the yield tables of \cite{Nomoto}, which also give an effective yield of $0.025$. This is a result of the wider SN progenitor mass coverage when combining both tables, which also results in more SN per stellar population.

The simulation also includes feedback from SNIa, AGB stars, and early stellar feedback. Additionally, the production and destruction of dust are modeled following the prescriptions in \citet{Aoyama2020,Romano_2022}. This comprehensive physics implementation enables the \textsc{Gadget4-Osaka} run to successfully meet the Cal-4 calibration criteria, as we discuss in the following sections.

\subsection{Physics Calibration Steps For \textsc{Gadget-4}}

In Section 5 of Paper III, we outlined a calibration procedure intended to minimize the number of free parameters in cross-code comparisons. This procedure consists of four steps: convergence in adiabatic gas evolution (Cal-1), implementation of the GRACKLE cooling--heating library (Cal-2), convergence of the stellar mass formed by $z=7$ in the absence of stellar feedback (Cal-3), and convergence to the stellar mass at $z=4$ once stellar feedback is included (Cal-4). In the following subsections, we present the outcome of each calibration step for \textsc{Gadget4-Osaka}. We will show \textsc{Gadget4-Osaka} right next to its predecessor \textsc{Gadget3-Osaka}, as it inherits many of its physics modules and code structure.

\subsubsection{Calibration Step One (\textsc{Cal-1}): Adiabatic Evolution of Gas}\label{sec:calstep1}

\begin{figure*}
    \centering
    \includegraphics[scale = 0.3]{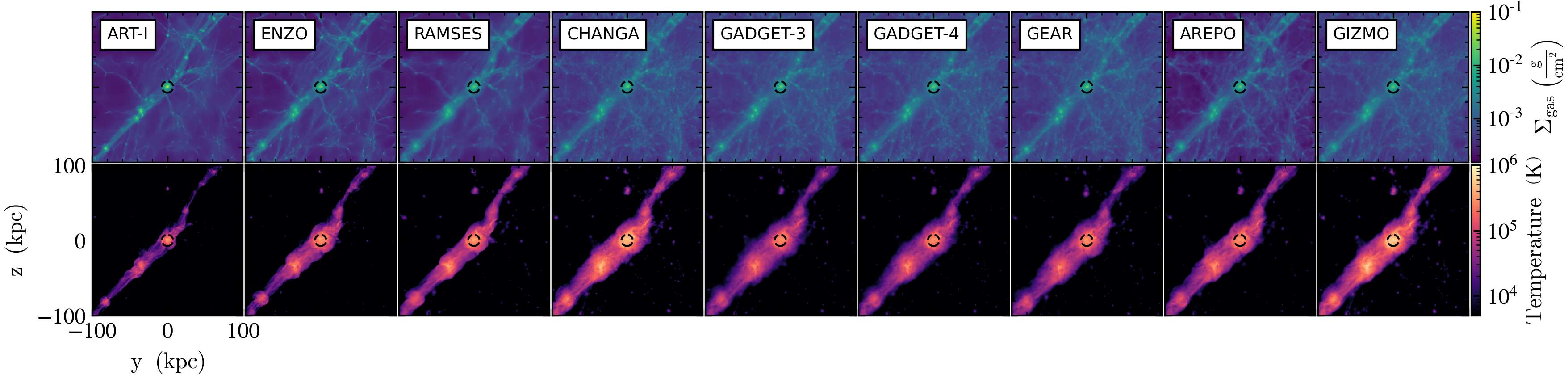}
    \caption{Gas density projection ({\it top}) and density-weighted temperature projection ({\it bottom}; each projected through a slab of thickness 200 kpc) at $z=7$ from the first calibration step, {\tt Cal-1} (adiabatic evolution test). 
        We indicate the mean $R_{\rm vir}$ among the codes ($\sim 7.5$ kpc at $z=7$) with a black dashed circle. 
        Units are proper kpc. 
        See Appendix \ref{sec:calstep1} for more information on {\tt Cal-1} and this figure.
	}
    \label{Ap:Cal1_0}
    \vspace{2mm}
\end{figure*}

In Figure~\ref{Ap:Cal1_0}, we present the projected density (top row) and temperature (bottom row) from the {\tt Cal-1} runs at $z=7$. The corresponding two-dimensional density--temperature probability distribution function (PDF) is shown in Figure~\ref{Ap:Cal1_1}. The large-scale density field and the multiphase structure in density--temperature space are again consistent across all nine codes, including the \textsc{Gadget4-Osaka} implementation. Minor variations arise primarily from differences in how each code handles hydrodynamic solvers and integration schemes. Further discussion on the origin of residual discrepancies can be found in Section 5.1 of Paper III.

\begin{figure*}
    \centering
    \begin{minipage}[t]{0.49\textwidth}
        \centering
        \includegraphics[scale = 0.125]{fig_G4/PDF_z7_Cal1.jpg}
        \caption{The $z=7$ composite of probability distribution function (PDF) of density and temperature for the gas within 100 kpc from the center of the main galactic system in the {\tt Cal-1} runs.  
            The 100 kpc-radius sphere encloses the main galaxy, the CGM, and the nearby IGM.  
            Colors represent the total gas mass in each 2-dimensional bin.  
            See Appendix \ref{sec:calstep1} for more information on {\tt Cal-1} and this figure.
            }
            \label{Ap:Cal1_1}
        \vspace{2mm}
    \end{minipage}
    \begin{minipage}[t]{0.49\textwidth}
        \centering
        \includegraphics[scale = 0.125]{fig_G4/PDF_z7_Cal2.jpg}
        \caption{The $z=7$ composite of 2-dimensional PDF of density and temperature for the gas within 100 kpc from the center of the main galactic system in the {\tt Cal-2} runs (cooling and heating test).  
            The 100 kpc-radius sphere encloses the main galaxy, the CGM, and the nearby IGM.  
            A black dashed vertical line is placed at the value of the star formation density threshold (see Section 3.1 of Paper III).
            See Appendix \ref{sec:calstep2} for more information on {\tt Cal-2} and this figure.
            }
            \label{Ap:Cal2_0}
        \vspace{2mm}
    \end{minipage}
\end{figure*}

\subsubsection{Calibration Step Two (\textsc{Cal-2}): Cooling and Heating of Gas By Common Physics Package}\label{sec:calstep2}

In Figure~\ref{Ap:Cal2_0} we show the two-dimensional density--temperature PDF from the {\tt Cal-2} runs (cooling and heating test) at 
$z=7$, comparing all nine participating codes. As in Figure 4 of Paper III, differences appear most clearly in the low-density high-temperature regime and in the high-density low-temperature gas. A detailed discussion of these variations can be found in Section 5.2 of Paper III. The \textsc{Gadget4-Osaka} {\tt Cal-2} run produces high-density low-temperature features consistent especially with the \textsc{Changa} and \textsc{Gizmo} runs. Aside from this, the results from \textsc{Gadget4-Osaka} fall comfortably within the expected code-to-code variations, confirming the correct implementation of the GRACKLE cooling–heating library.

\subsubsection{Calibration Step Three (\textsc{Cal-3}): Common Star Formation Physics}\label{sec:calstep3}

Figure~\ref{Ap:Cal3_0} shows the composite two-dimensional density–temperature PDF at $z=7$ from the {\tt Cal-3} runs (star formation test). As discussed in Section 5.3 of Paper III, there is overall agreement in the temperature–density features across all codes, now also including the \textsc{Gadget4-Osaka} run. In particular, \textsc{Gadget4-Osaka} produces results that closely track those of other SPH-based codes, demonstrating consistency in the treatment of star-forming gas. The success of the {\tt Cal-3} calibration is highlighted in Figure~\ref{Ap:Cal3_1}, where all code groups converge to a final stellar mass of $\sim 2\times10^9\, {\rm M}_{\odot}$ at $z=7$, within 0.35 dex. Section 5.3 of Paper III contains a detailed discussion of differences between codes. Finally, Figure~\ref{Ap:Cal3_2} presents the projected gas density (top row), temperature (middle row), and stellar surface density (bottom row) at $z=7$. The \textsc{Gadget4-Osaka} results show good convergence with the other codes in both density and temperature distributions, with only minor offsets in the gas morphology and temperature projections, already noted in Section 5.3.2 of Paper III.

\begin{figure*}
    \centering
    \includegraphics[scale = 0.125]{fig_G4/PDF_z7_Cal3.jpg}
    \caption{The $z=7$ composite of 2-dimensional PDF of density and temperature for the gas within 100 kpc from the center of the main galactic system in the {\tt Cal-3} runs (star formation test).  
        The 100 kpc-radius sphere encloses the main galaxy, the CGM, and the nearby IGM.  
        A black dashed vertical line marks the density threshold for star formation.
        See Appendix \ref{sec:calstep3} for more information on {\tt Cal-3} and this figure.        
        }
        \label{Ap:Cal3_0}
    \vspace{2mm}
\end{figure*}

\begin{figure*}
    \centering
    \includegraphics[scale = 0.32]{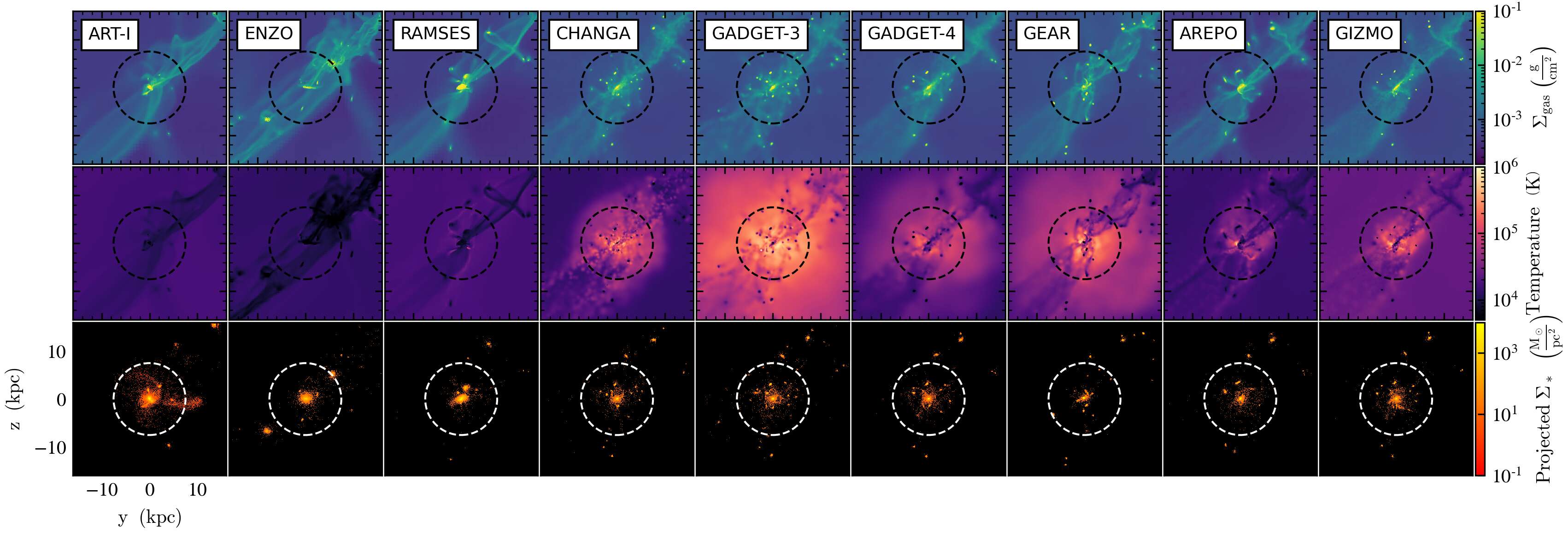}
    \caption{Gas density projection ({\it top}), density-weighted temperature projection ({\it middle}), and stellar surface density ({\it bottom}) at $z=7$ from the third calibration step, {\tt Cal-3}. 
        The width of each panel is $4R_{\rm vir} = 30 \,{\rm kpc}$. 
        The mean $R_{\rm vir}$ among the codes ($\sim 7.5$ kpc) is indicated with a black/white dashed circle.
        See Appendix \ref{sec:calstep3} for more information on {\tt Cal-3} and this figure.            
	}
        \label{Ap:Cal3_2}
    \vspace{2mm}
\end{figure*}

\subsubsection{Calibration Step Four (\textsc{Cal-4}): “Favorite” Stellar Feedback Prescription By Each Code}\label{sec:calstep4}

\begin{figure*}
    \centering
    \begin{minipage}[t]{0.49\textwidth}
        \centering
        \includegraphics[scale = 0.2375]{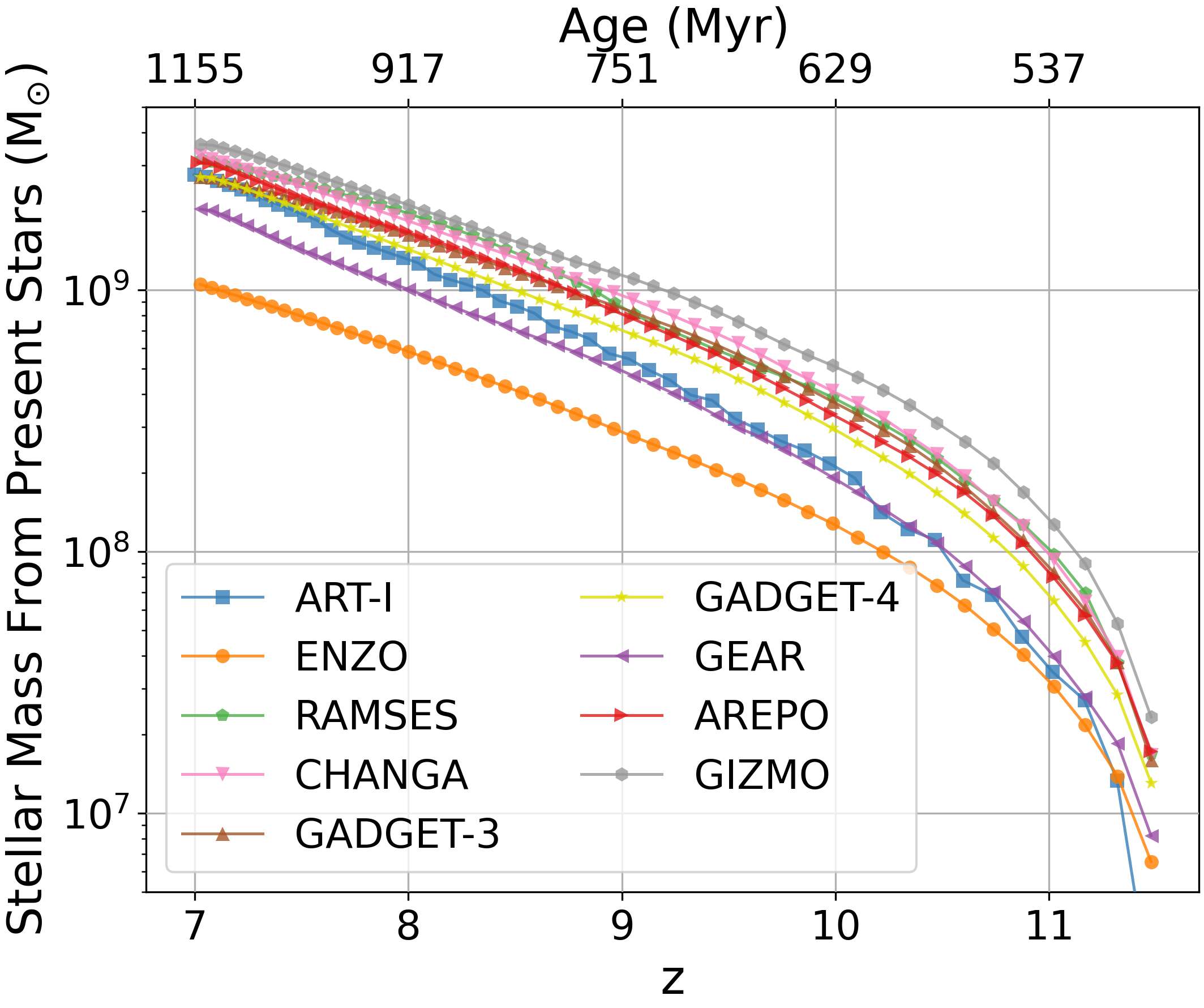}
       \caption{Stellar mass growth histories for the {\tt Cal-3} runs in a 100 kpc sphere centered at the target progenitor. The curve is computed using the ages or creation times recorded in star particles at $z=7$. See Appendix \ref{sec:calstep3} for more information on {\tt Cal-3} and this figure.           
            }
            \label{Ap:Cal3_1}
        \vspace{2mm}
    \end{minipage}
    \begin{minipage}[t]{0.49\textwidth}
        \centering
        \includegraphics[scale = 0.235]{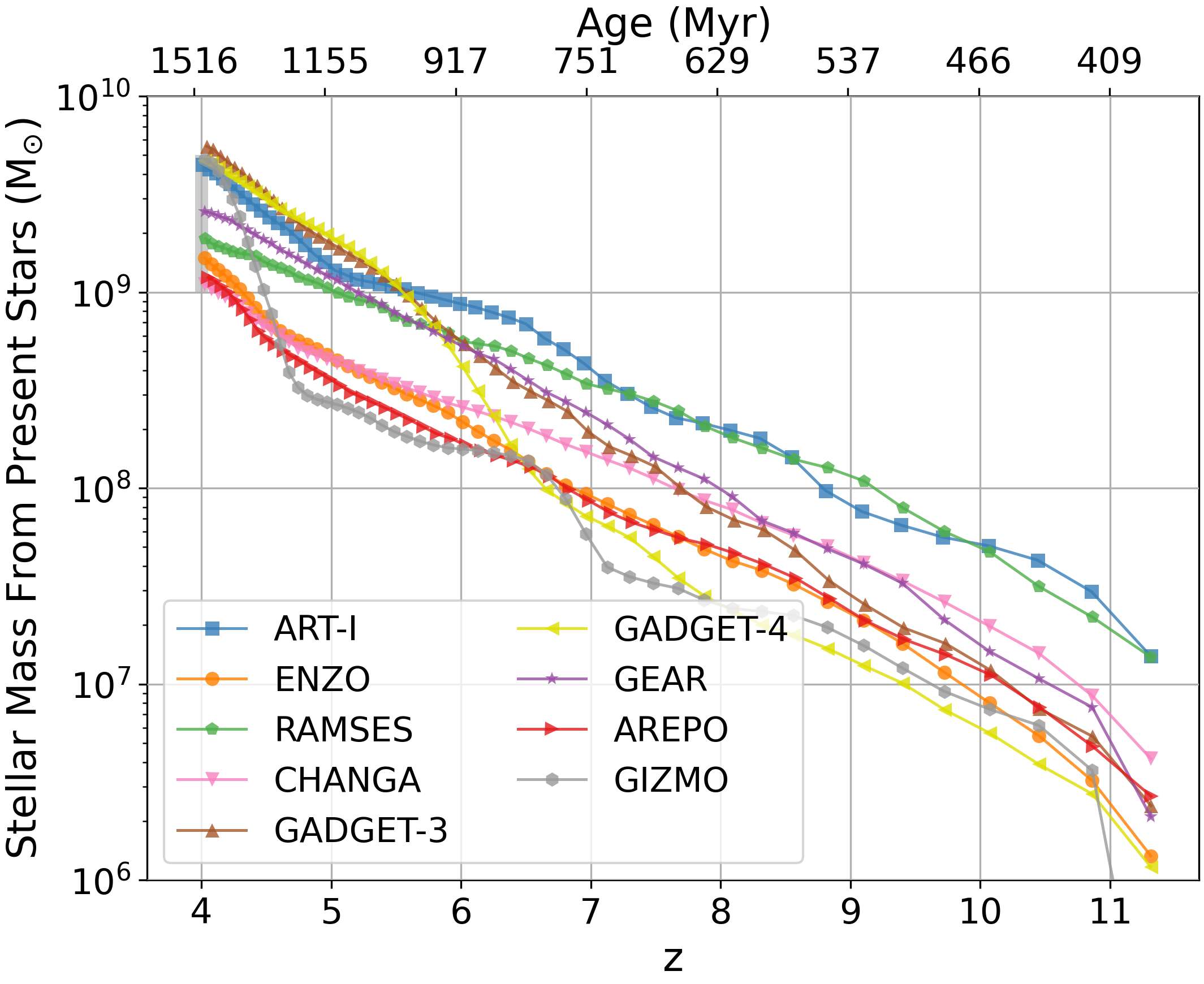}
        \caption{Stellar mass growth histories for the {\tt Cal-4} runs inside a $R_{\rm vir}$ sphere centered at the target progenitor. The curve is computed using the ages or creation times recorded in star particles at $z=4$. Therefore, what we show here is an upper limit for the total $M_\star$ formed inside $R_{\rm vir}$. The stellar mass range at $z=4$ targeted in our calibration is $M_\star \sim 1-5\times 10^9\, {\rm M}_{\odot}$, as motivated by semi-empirical models. See Appendix \ref{sec:propertiesz4} for more information on \textit{CosmoRun} and this figure.   }
        \label{Ap:Cal4_0}
        \vspace{2mm}
    \end{minipage}
\end{figure*}

In this final calibration step, {\tt Cal-4}, we compare the cosmological simulations obtained with each group’s fiducial stellar feedback model, now also including the \textsc{Gadget4-Osaka} simulation. Figure \ref{Ap:Cal4_0} shows the convergence of the stellar mass at $z=4$ toward the predictions of semi-empirical models. The stellar mass growth in the \textsc{Gadget4-Osaka} run closely follows that of the other codes, remaining within the overall scatter at all times. This confirms that the updated feedback implementation in \textsc{Gadget4-Osaka} is well aligned with the calibration goals. With this {\tt Cal-4} result, we complete the calibration procedure for \textsc{Gadget4-Osaka}. The nine simulations shown in Figure~\ref{Ap:Cal4_0} together form the updated \textit{CosmoRun} suite. In the next section, we turn to an analysis of these simulations, focusing on the \textsc{Gadget4-Osaka} result at $z=4$.

\subsubsection{Global Properties of The Target Galaxy Progenitor In The \textsc{Gadget-4} CosmoRun At z = 4}\label{sec:propertiesz4}

In this section, we reproduce a set of comparison plots resembling those in Section 6 of Paper III, and equal to those presented for \textsc{Arepo-t} in Appendix A of Paper IV. As found in our earlier analysis, overall bulk properties tend to converge between codes, while morphological details, temperature structure, and metal distribution remain sensitive to each group’s stellar-feedback implementation.

Figure \ref{Ap:Cal4_1} displays the two-dimensional density–temperature PDF for gas within the codes’ mean $R_{\rm vir}$ ($\sim 25.4$ kpc) at $z=4$. The \textsc{Gadget4-Osaka} run matches the general temperature--density trends seen in the other participants, and shows especially close agreement with \textsc{Gizmo} and \textsc{Arepo-t}. Figure \ref{Ap:Cal4_2} then compares the spatial distributions of dark matter, stars and gas, together with maps of gas temperature and metallicity for all nine codes at around $z \sim 4$. Large-scale structure around the target halo is consistent across the suite, but the maps reveal code-dependent variation in galactic morphology, gas temperature, and metal enrichment. These differences mostly trace back to distinct feedback prescriptions and yield choices, as previous calibration steps demonstrate. In particular, the overall higher metallicity seen in \textsc{Gadget4-Osaka} arises due to the yield tables, as we discussed in Section \ref{sec:g4phys}. Timing differences analyzed in Appendix C of Paper IV can also contribute to some of the offsets between runs.

Figures \ref{Ap:Cal4_6} and \ref{Ap:Cal4_3} to \ref{Ap:Cal4_5} examine these gas properties in more detail. In particular, Figure \ref{Ap:Cal4_6} again shows the two-dimensional density–temperature PDF, now colored by metal mass per bin, and indicates that the \textsc{Gadget-4} run exhibits a slightly higher metal enrichment than other codes in most of the gas. Figure \ref{Ap:Cal4_3} shows that the spherically averaged gas density profiles converge well among the codes. Figures \ref{Ap:Cal4_4} and \ref{Ap:Cal4_5} demonstrate that inter-code differences in metal enrichment are most pronounced in the highest and lowest-metallicity bins; these extremes reflect the combined effect of feedback strength, diffusion scheme, and adopted yields. Because gas particles inherit the metallicity of star particles, similar inter-code distinctions appear in the stellar metallicity distribution shown in Figure \ref{Ap:Cal4_5}.

Taken together, these calibrated comparisons indicate that the \textsc{Gadget4-Osaka} implementation is consistent with the trends reported in Paper III. With \textsc{Gadget4-Osaka} included, the \textit{CosmoRun} suite now comprises nine simulations.

\begin{figure*}
    \centering
    \begin{minipage}[t]{0.49\textwidth}
        \centering
        \includegraphics[scale = 0.125]{fig_G4/PDF_z4_Cal4.jpg}
        \caption{The $z=4$ composite of 2-dimensional PDF of density and temperature for the gas within the mean $R_{\rm vir}$ among the codes  ($\sim 25.4$ kpc)  from the target galaxy's center in the \textit{CosmoRun} simulations. A black dashed vertical line marks the density threshold for star formation. See Appendix \ref{sec:propertiesz4} for more information on \textit{CosmoRun} and this figure.                 
            }
        \label{Ap:Cal4_1}
        \vspace{2mm}
    \end{minipage}
    \begin{minipage}[t]{0.49\textwidth}
        \centering
        \includegraphics[scale = 0.125]{fig_G4/metal_PDF_z4_Cal4.jpg}
        \caption{Similar to Figure~\ref{Ap:Cal4_1}, but now with colors representing the total metal mass in each 2-dimensional bin in our \textit{CosmoRun} simulation suite. Note that the PDF is for the gas within $R_{\rm vir}$ from the center of the target galaxy in the \textit{CosmoRun} simulations. A sphere of radius $R_{\rm vir}$ encloses the main galaxy and CGM, but not the IGM. See Appendix \ref{sec:propertiesz4} for more information on  \textit{CosmoRun} and this figure.
            }  
            \label{Ap:Cal4_6}
        \vspace{2mm}
    \end{minipage}
\end{figure*}

\begin{figure*}
    \centering
    \includegraphics[scale = 0.33]{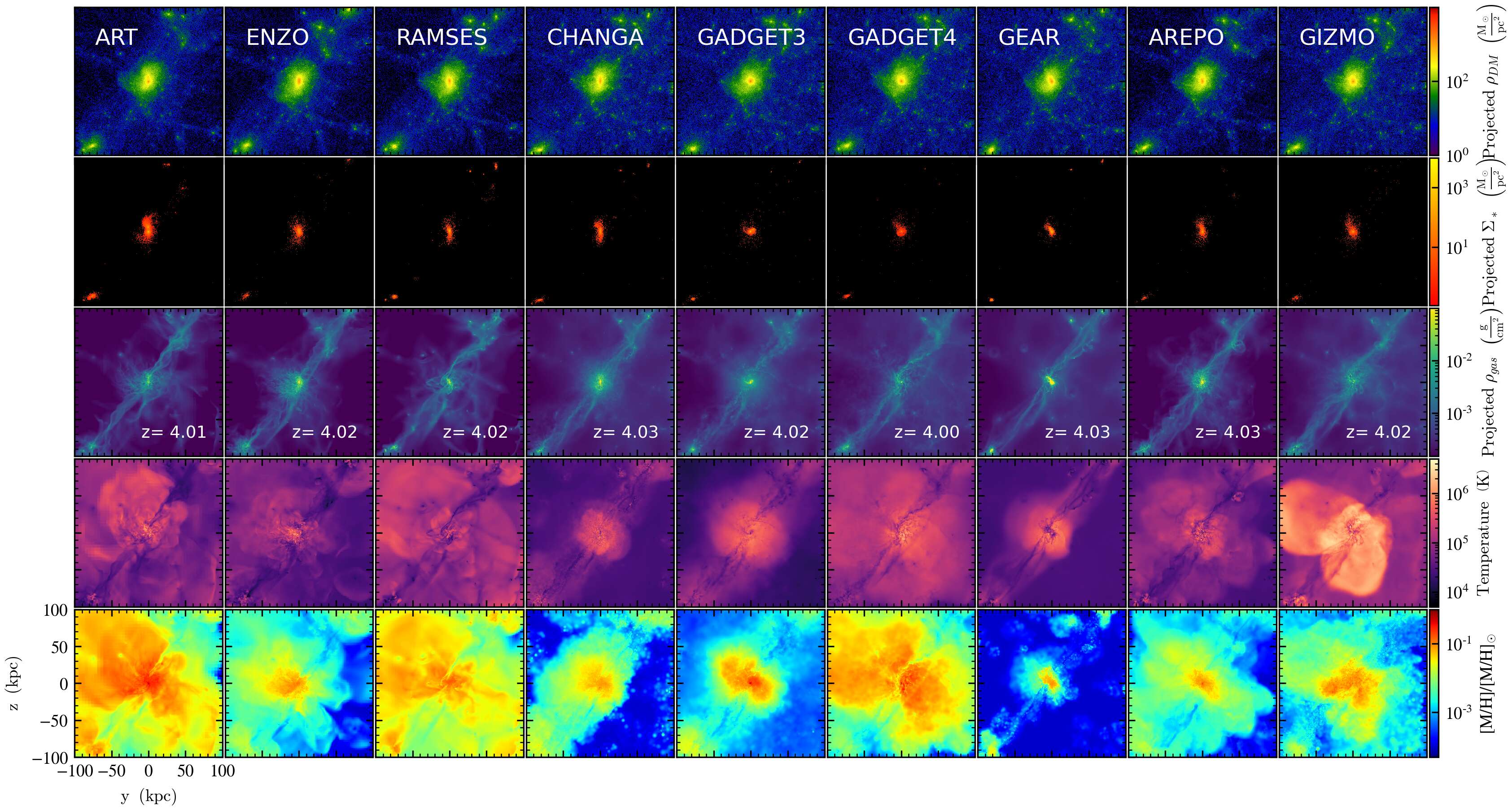}
    \caption{Dark matter surface density ({\it 1st row}), stellar surface density ({\it 2nd row}), gas surface density ({\it 3rd row}), density-weighted projection of gas temperature ({\it 4th row}), and density-weighted projection of gas metallicity ({\it 5th row}) through a slab of thickness 200 kpc at $z=4$ in our \textit{CosmoRun} simulation suite. 
    See Appendix \ref{sec:propertiesz4} for more information on \textit{CosmoRun} and this figure. Simulations performed by:  Santi Roca-F\`{a}brega (\textsc{Art-I}, \textsc{Ramses}), Ji-hoon Kim (\textsc{Enzo}), Johnny Powell and H\'ector Vel\'azquez (\textsc{Changa}), Kentaro Nagamine and Ikkoh Shimizu (\textsc{Gadget-3}), Kentaro Nagamine and Pablo Granizo (\textsc{Gadget-4}), Loic Hausammann and Yves Revaz (\textsc{Gear}), Anna Genina (\textsc{Arepo-t}), and Alessandro Lupi and Bili Dong (\textsc{Gizmo})}
    \label{Ap:Cal4_2}
    \vspace{2mm}
\end{figure*}

\begin{figure*}
    \centering
    \includegraphics[scale = 0.23]{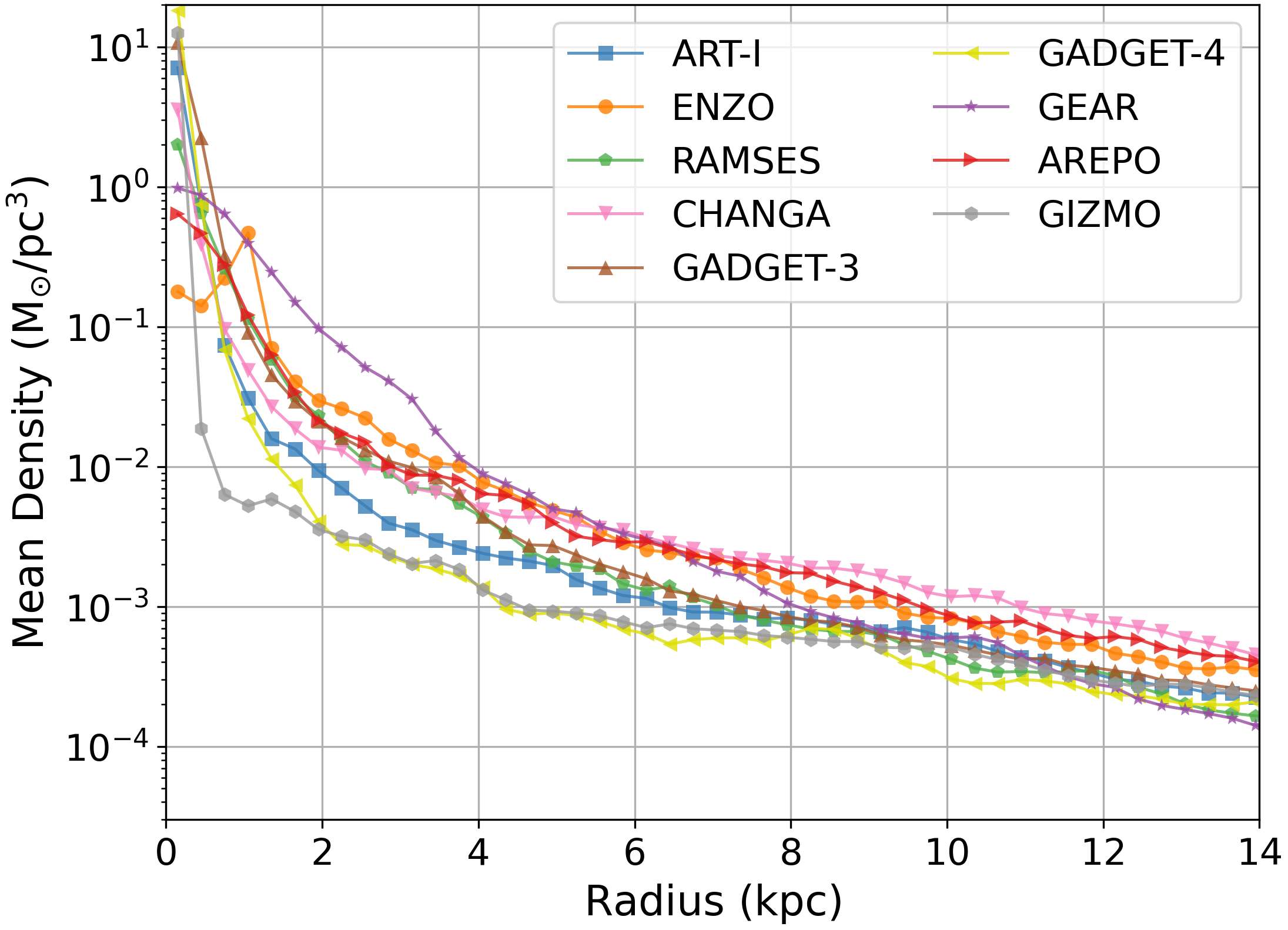}
   \caption{Spherically-averaged gas density profile as a function of distance from the target progenitor's center at $z=4$ in our \textit{CosmoRun} simulation suite. See Appendix \ref{sec:propertiesz4} for more information on \textit{CosmoRun} and this figure.    
        }
        \label{Ap:Cal4_3}
    \vspace{2mm}
\end{figure*}

\begin{figure*}
    \centering
    \begin{minipage}[t]{0.49\textwidth}
        \centering
        \includegraphics[scale = 0.2325]{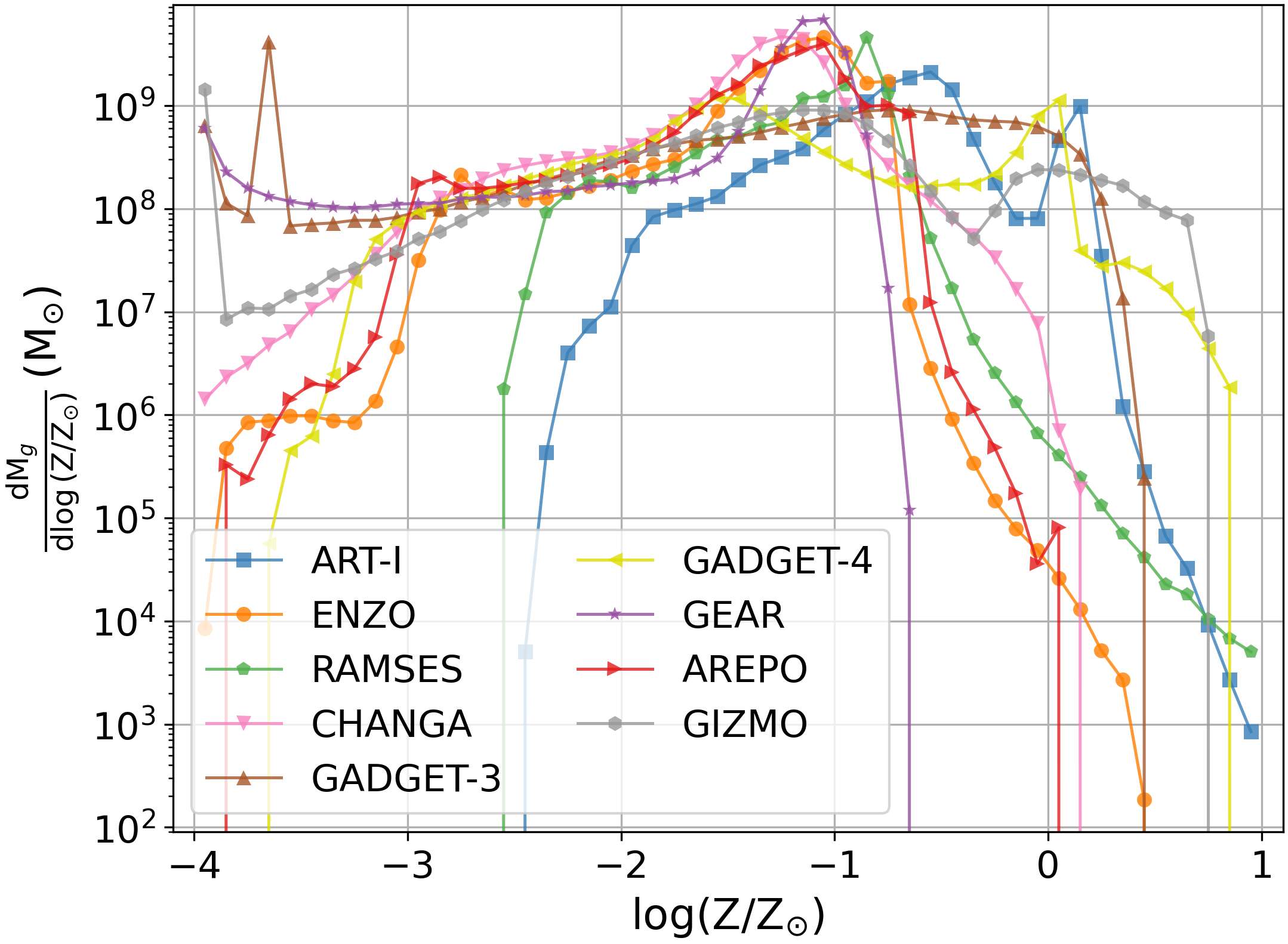}
        \caption{Distribution of gas mass as a function of gas metallicity at $z=4$ for all the gas inside the target progenitor's $R_{\rm vir}$ in our \textit{CosmoRun} simulation suite. See Appendix \ref{sec:propertiesz4} for more information on \textit{CosmoRun} and this figure. 	
            }
            \label{Ap:Cal4_4}
        \vspace{2mm}
    \end{minipage}
    \begin{minipage}[t]{0.49\textwidth}
        \centering
        \includegraphics[scale = 0.2275]{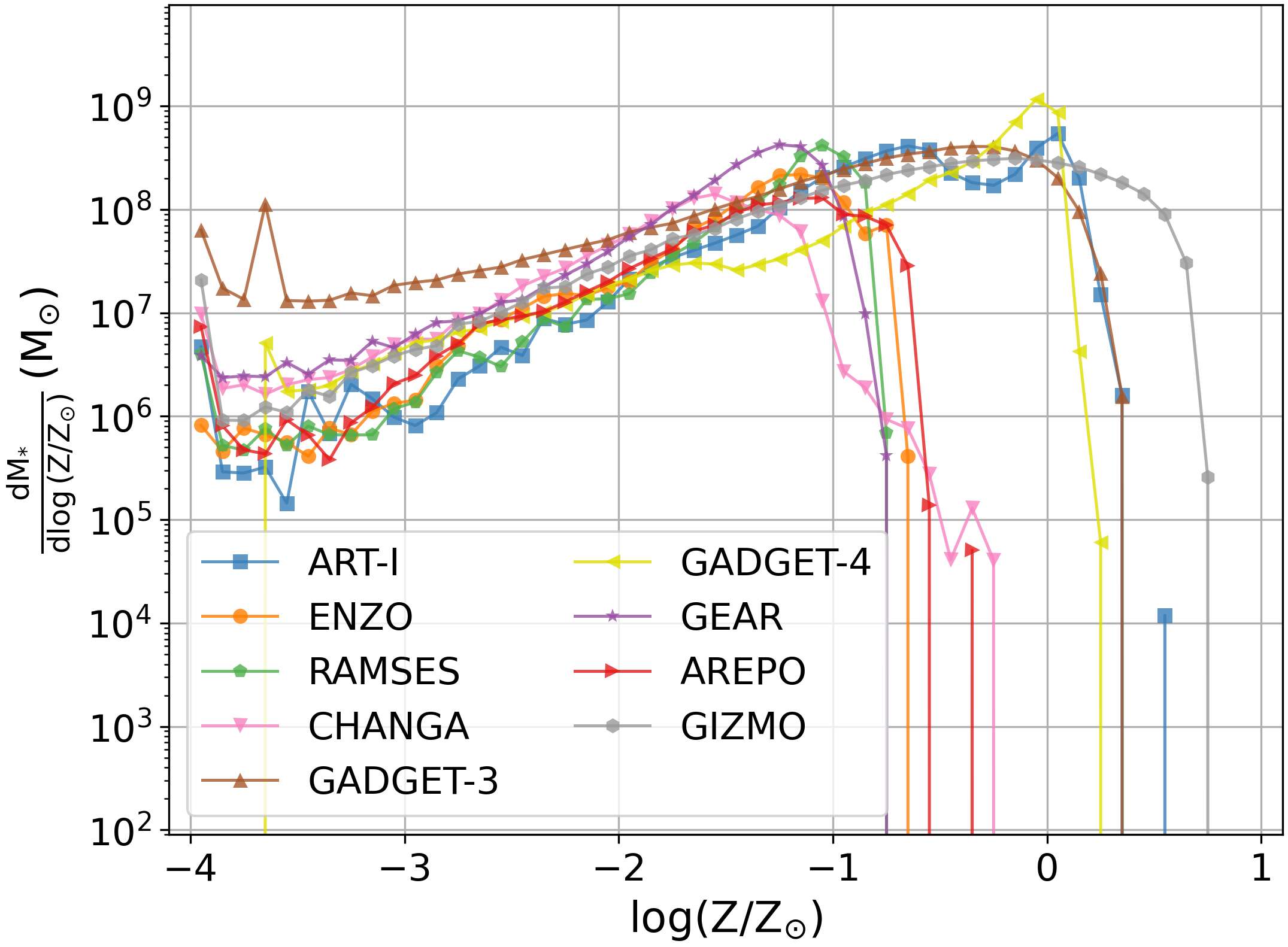}
        \caption{Distribution of stellar mass as a function of stellar metallicity at $z=4$ for all the stars inside the target progenitor's $R_{\rm vir}$ in our \textit{CosmoRun} simulation suite. See Appendix \ref{sec:propertiesz4} for more information on \textit{CosmoRun} and this figure.}
            \label{Ap:Cal4_5}
        \vspace{2mm}
    \end{minipage}
\end{figure*}



\end{document}